\documentclass{aa}  
\usepackage{graphicx}
\usepackage{txfonts}
\begin{document}

   \title{AMICO galaxy clusters in KiDS-DR3: Evolution of the luminosity function between z=0.1 and z=0.8}


   \author{E. Puddu \inst{1}, 
          M. Radovich \inst{2},
          M. Sereno \inst{3,7},
          S. Bardelli \inst{3},
          M. Maturi  \inst{5,6},
          L. Moscardini \inst{4,3,7},
          F. Bellagamba \inst{4,3,7},
          C. Giocoli \inst{4,3,7}, 
          F. Marulli \inst{4,3,7},
          M. Roncarelli \inst{4,3,7}
          }

   \institute{I.N.A.F., Osservatorio Astronomico di Capodimonte,
              Salita Moiariello 16, Napoli 80131, Italy\\
              \email{emanuella.puddu@inaf.it}
         \and
             I.N.A.F., Osservatorio Astronomico di Padova, 
             Vicolo dell'Osservatorio 5, Padova 35122, Italy
         \and
             I.N.A.F., Osservatorio di Astrofisica e Scienza dello Spazio di Bologna, 
             Via Gobetti 93/3, I-40129 Bologna, Italy
          \and
             Dipartimento di Fisica e Astronomia, Alma Mater Studiorum Universit\`a di Bologna, 
             Via Gobetti 93/2, I-40129 Bologna, Italy
           \and
            Zentrum f\"ur Astronomie, Universit\"at Heidelberg, Philosophenweg 12, D-69120 Heidelberg, Germany
           \and
            ITP, Universit\"at Heidelberg, Philosophenweg 16, 69120 Heidelberg, Germany
           \and
            I.N.F.N. - Sezione di Bologna, 
            Viale Berti Pichat 6/2, I-40127 Bologna, Italy
            }

   \date{Received Month Day, Year; accepted Month Day, Year}

 
\abstract
   {}
   {By means of the $r$-band luminosity function  (LF)  of galaxies in a sample of about 4000 clusters detected by the cluster finder AMICO in the KiDS-DR3 area of about 400~deg$^2$, we studied the evolution with richness and redshift of the passive evolving (red), star-forming (blue), and total galaxy populations. This analysis was performed for clusters in the redshift range [0.1,0.8] and in the mass range [$10^{13} M_{\odot}$,$10^{15} M_{\odot}$].}
   {To compute LFs, we binned the luminosity distribution in magnitude and statistically subtracted the background. Then, we divided the cluster sample in bins of both redshift and richness/mass. 
   We stacked LF counts in each 2D bin for the total, red, and  blue  galaxy  populations; finally, we  fitted  the stacked LF with a Schechter function and studied the trend of its parameters with redshift and richness/mass.}
   {We found a passive evolution with $z$ for the bright part of the LF for the red and total populations and no significant trends for the faint galaxies. The  mass/richness dependence is clear for the density parameter $\Phi_{\star}$, increasing with richness, and for the total population faint end, which is shallower in the rich clusters.}
   {}


    \titlerunning{Cluster luminosity function in AMICO-KiDS}
    \authorrunning{Puddu et al.}
   \maketitle
   
%

\section{Introduction}

One of the fundamental constraints on the different scenarios of galaxy evolution and formation is how  the luminosity function (LF) of galaxies in clusters changes with redshift and environment. The LF is defined as the number density of galaxies per absolute magnitude as a function of luminosity \citep{Peebles1971}; the observed LFs can be well described by a Schechter function \citep{Schech1976} as follows:
\begin{equation} \label{Schechter}
    \Phi(M)=\Phi_{\star} 10^{0.4 (\alpha+1) (M-M_{\star})} exp[-10^{0.4 (M-M_{\star})}] .
\end{equation}
This function is parametrized by a characteristic amplitude $\Phi_{\star}$ and magnitude M$_{\star}$, and by a faint-end slope $\alpha$, each of those parameters being related to some cluster property or physical process relevant to galaxy evolution.

Studies of the LFs animated the debate about the deficit of faint red galaxies with increasing redshift \citep{DeLucia2004,DeLucia2007,Goto2005,Tanaka2005}, which has implications for the two main evolutionary scenarios: the hierarchical 
and the passive evolution 
model. 
In the case of the red galaxies deficit (supporting the first evolutionary scenario), the LF faint-end slope $\alpha$ parameter becomes shallower with $z$ for the red-sequence galaxies
 \citep[as suggested by][]{DeLucia2007,Stott2007,Gilbank2008,Rudnick2009,DeFilippis2011,Martinet2015,Lin2017}. If $\alpha$ remains constant  \citep[as found by][]{dePropris2007,dePropris2013,dePropris2015,dePropris2016,Andreon2006,Andreon2008,Strazzullo2006,Strazzullo2010,Mancone2012,Wylezalek2014,Cerulo2016,Sarron}, then the second scenario is favoured.
The result that $M_{\star}$ evolves passively out to about $z=$1.5  \citep{Strazzullo2010,Mancone2010,Andreon2013,Andreon2014,Wylezalek2014} 
indicates that, at least  for bright cluster galaxies, the latter is the preferred scenario.

The disagreement among these studies and results is hard to interpret given the differences in the cluster samples and their selection function, the presence of cluster-to-cluster LF variations, surface brightness selection effects on galaxies \citep{dePropris2013}, and even dissimilarities in the methods adopted to compute single cluster or stacked LFs. Moreover, the LF evolution in cluster environments depends on properties relative to the whole cluster, such as mass \citep{Gilbank2008,Hansen2009,2016Lan} and dynamical state \citep{WH2015,dePropris2013,Zenteno2020}, as well as on properties of member galaxies, such as galaxy types (star-forming or passive) \citep{Goto2002,Muzzin2007}, environment, and density \citep{dePropris2003,Hansen2005,Lanzoni2005,Popesso2006,DeFilippis2011}.

Our present understanding of galaxy evolution arises from the comparison between local large cluster samples detected optically \citep[e.g. Sloan Digital Sky Survey;][]{sdss2000} or in X-rays \citep[ROSAT All-Sky Survey;][]{rosat11998, rosat2000,rosat32004} and high-redshift cluster samples derived from surveys of few square degrees, sometimes using infrared data.
More recently, the Sunyaev-Zel'dovich effect \citep[][]{SZ1970} has been used in wide surveys, such as those carried out with the South Pole Telescope \citep[SPT-SZ;][]{SPTsurvey2015} and with Planck \citep{Planck2014}, to enlarge the sample of distant and massive clusters \citep{Zenteno2016}. Moreover, ongoing optical surveys, for example the Dark Energy Survey \citep[DES;][]{DES2016}, the Hyper Suprime-Cam Subaru Strategic Program \citep[HSC-SSP;][]{HSCSPP2018}, and the just-completed Kilo Degrees Survey \citep[KiDS;][]{deJong2017} have begun to provide cluster samples in wide ranges of redshift and mass \citep{Maturi,Hennig} enabling more comprehensive studies. Future wide-field optical surveys planned on the new projects Euclid \citep{Euclid} and Large Synoptic Survey Telescope \citep[LSST;][]{collaboration2009lsst} will be crucial to settle the discrepancies in the observations and to validate one of the two evolutive scenarios.

A very important role in the study of galaxy clusters is played by the cluster detection algorithm. 
The most popular method at optical or infrared wavelengths, efficient at low ($z<$0.3) and intermediate (0.3$>z>$0.6) redshifts, is based on the presence of the red sequence \citep{redmapper}: this method has been used for SDSS, DES, and for the  HSC-SSP Survey \citep[Camira;][]{CAMIRA2014}. At higher redshift, where the red sequence is expected to be less prominent, it could be useful to adopt criteria not based on colour selection such as searching for galaxy over-densities or using  weak-lensing convergence maps of background galaxies \citep{convmaps2018,convmaps2020}. 
In this work, we used a linear optimal matched filter algorithm, which takes advantage of the photometric redshifts and the magnitude of the galaxies. In this way, forming clusters, which have a relevant fraction of blue young galaxies and whose red sequence is still not well defined, are not penalized by an explicit colour selection of the galaxy population. 

In this paper we study the evolution of the $r$-band LF of the galaxies belonging to the AMICO-KiDS-DR3 cluster sample \citep{Maturi} for the total, the early-type (ET), and late-type (LT) galaxy populations as a function of cluster richness/mass and redshift.
The structure of the paper is as follows: We first describe the cluster catalogue, how it was derived from the KiDS-DR3 data, and the outcomes (Sect.~\ref{clcatderi}).
In Sect.~\ref{clusterLF} we describe the methods used to construct and parametrize the single and the composite LFs and we discuss the red and blue galaxies selection.
The results are given in Sect.~\ref{result}, compared with literature in Sect.~\ref{coparison_lit}, discussed in Sect.~\ref{discussion}, and the conclusions are drawn in Sect.~\ref{conclusion}.
Throughout the paper, we assume a $\Lambda$CDM model with $\Omega_M=$ 0.3, $\Omega_{\Lambda}=$ 0.7, and H$_0=$ 70 km Mpc$^{-1}$s$^{-1}$.

\section{Cluster sample: Catalogue derivation and features}
\label{clcatderi}

In our study we consider data coming from KiDS, which is an ESO Public Survey in the $ugri$ bands (with limiting magnitudes of 24.3, 25.1, 24.9, 23.8, respectively) carried out with the  Very Large Telescope Survey Telescope (VST) and the OmegaCAM wide-field camera. The KiDS survey comprises 1350 deg$^2$ divided between two regions: an equatorial stripe (KiDS-N) and a south  Galactic  pole  stripe (KiDS-S). At the moment, available public data releases are KiDS-DR2 \citep[$\sim$~100 deg$^2$;][]{DEJONG2015}, KiDS-DR3 \citep[$\sim$~440 deg$^2$;][]{deJong2017}, and KiDS-DR4 \citep[$\sim$~1000 deg$^2$;][]{Kui2019}, the latter including $ZYJHK_s$ photometry from the VIKING survey on the VISTA telescope. 

The cluster catalogue used in this work \citep{Maturi} was derived from the KiDS-DR3 photometric data on which we run the Adaptive Matched Identifier of Clustered Objects \citep[hereafter AMICO;][]{Bellag18}. We selected
AMICO to be implemented in the Euclid pipeline as the result of the Euclid Cluster Finder Challenge \citep{Euclid} and AMICO was first applied to KiDS data on the   $\sim$~100~deg$^2$ DR2 area \citep{paper1}. 
The AMICO-KiDS-DR3 catalogue, considered in this work, contains 7988 clusters in the redshift range 0.1$<z<$~0.8 detected in $\sim$~400~deg$^2$  with a signal-to-noise ratio $S/N\geq$~3.5. 

The colour properties of this cluster sample were investigated in \cite{MarioBCG}. These authors define how to select the brightest central galaxy (hereafter BCG); then, they define the criteria to classify blue and red cluster members. Finally, they statistically analyse the red/blue fraction of the brightness and stellar mass of the central galaxy and the magnitude gap as a function of redshift and cluster mass.

As previously mentioned, AMICO was built on a cluster model, which takes into account the luminosity and spatial distribution of galaxies, but disregards any colour selections. The a priori radial profile is modelled as a Navarro-Frenk-White \citep[NFW;][]{NFW97}. The prior model for the LF is the Schechter function with characteristic parameters from \citet{Hennig}: these parameters were derived from a Sunyaev Zel'dovich selected clusters sample observed by DES \citep{Hennig,Zenteno2016}. In particular, regarding the LF parameters, \citet{Hennig} found that the shape of the characteristic magnitude evolution with redshift, $m_{\star}(z)$, is compatible with a stellar population evolutionary model with a decaying starburst at redshift $z = 3$ (decay time $= 0.4$ Gyr) and a Chabrier initial mass function \citep{B&C03}; they also found a mean faint-end slope $\alpha$ of $-1.06$. 

\citet{Maturi}, employing mock galaxy catalogues based on the KiDS-DR3 data, derived the sample purity and completeness for the AMICO cluster catalogue: the purity approaches 95$\%$ over almost the whole richness and redshift range \citep[see Fig.~12 of][]{Maturi}, and the completeness is larger than 80$\%$ at low/intermediate redshift and for high/intermediate richness.  
The AMICO cluster catalogue lists, among other properties, position, redshift, and the mass proxy $\lambda_{\star}$ \citep[for a complete list, see Table 3 of][]{Maturi}.

The intrinsic richness $\lambda_{\star}$ is derived by summing the probabilities over the galaxies brighter than the characteristic magnitude $m_{\star}(z) + 1.5$ and within a fixed radius, defined as the $\tilde{r}_{200}$ for a typical cluster with a mass of $\tilde{M}_{200}$=10$^{14} M_{\odot}/h$; $m_{\star}(z)$ is the same used in the LF model. 
This richness definition samples a radial distance from the cluster centre that is large enough to encompass a comprehensive fraction of galaxies in a broad mass range. It also samples, for each cluster, the same portion of the LF that is faint enough not to be affected by incompleteness; this ensures that $\lambda_{\star}$ is a redshift independent mass proxy. 

\citet{Bellag19} performed a weak-lensing stacked analysis by binning the cluster catalogue in redshift and $\lambda_{\star}$. In this way, these authors provide a calibration of the relation between the cluster mass $M_{200}$ and the mass proxy $\lambda_{\star}$, which is fundamental for astrophysical and cosmological studies (Lesci et al., in preparation; Nanni et al., in preparation).  Each candidate cluster falling in the redshift range 0.1~$<z<$~0.6 has an associate mass estimated from richness; for higher redshift the scaling relation has to be used as an extrapolation.
Thanks to the AMICO performances, the calibration extends to low-mass groups, down to $M_{200}$ $\sim$~2$\times 10^{13} M_{\odot}/h$ at $z=0.2$ and $M_{200}$ $\sim$ 5$\times 10^{13} M_{\odot}/h$ at $z=0.5$ \citep{Bellag19}.

\section{Cluster galaxy luminosity functions}
\label{clusterLF}

\subsection{Luminosity function by statistical subtraction}
\label{LFbySS}

To determine the cluster LF, we use the $r$ band because it is deeper and has better seeing than the other filters.
The main step is the selection of the member galaxies and the determination of their distribution in luminosity. 
 The AMICO cluster finder returns a probabilistic association of galaxies to each cluster; in particular it produces, for each galaxy $i$, the probability $P_i(j)$ of belonging to the $j$th cluster detection. But we decided not to use it because it is model dependent.

To perform our analysis, we chose the approach for which the background contribution to the galaxies counts, locally estimated in an annulus around the cluster, is statistically subtracted by the number of galaxies within a projected radial distance $r_{in}$ from the cluster centre. 
This method allows us to take into account the field non-uniformity of the projected galaxy density at cluster scales.
The background annulus should be close enough to the cluster over-density to be representative of the field galaxies distribution along the cluster line of sight. At the same time, contaminations by the cluster galaxies, which dilute the cluster signal \citep[see][]{Paolillo2001}, should be avoided by choosing the annulus not too close to the central region.
Therefore, we define the background annulus from $2$ to $3\cdot r_{in}$ (see Fig.~\ref{Fig1}). 

Since our cluster catalogue was obtained by running AMICO on the single tiles of the KiDS-DR3, a correct estimate of the background is possible only for those over-densities lying not too near to the borders, so that the background region is completely included in the belonging tile. For this reason, we excluded those clusters whose distance from the tile edge is less than the outer radius of the background annulus.

We also removed all the clusters affected by areas masked because of image artefacts; the parameter MASKFRAC in the cluster catalogue  \citep[see Table 3 of][]{Maturi} indicates the masked fraction of the detection area, and we excluded the detections with MASKFRAC $\geq 0.1$ (i.e. a masked area of $< 10\%)$.

\begin{figure}
   \centering
   \includegraphics[width=9.cm]{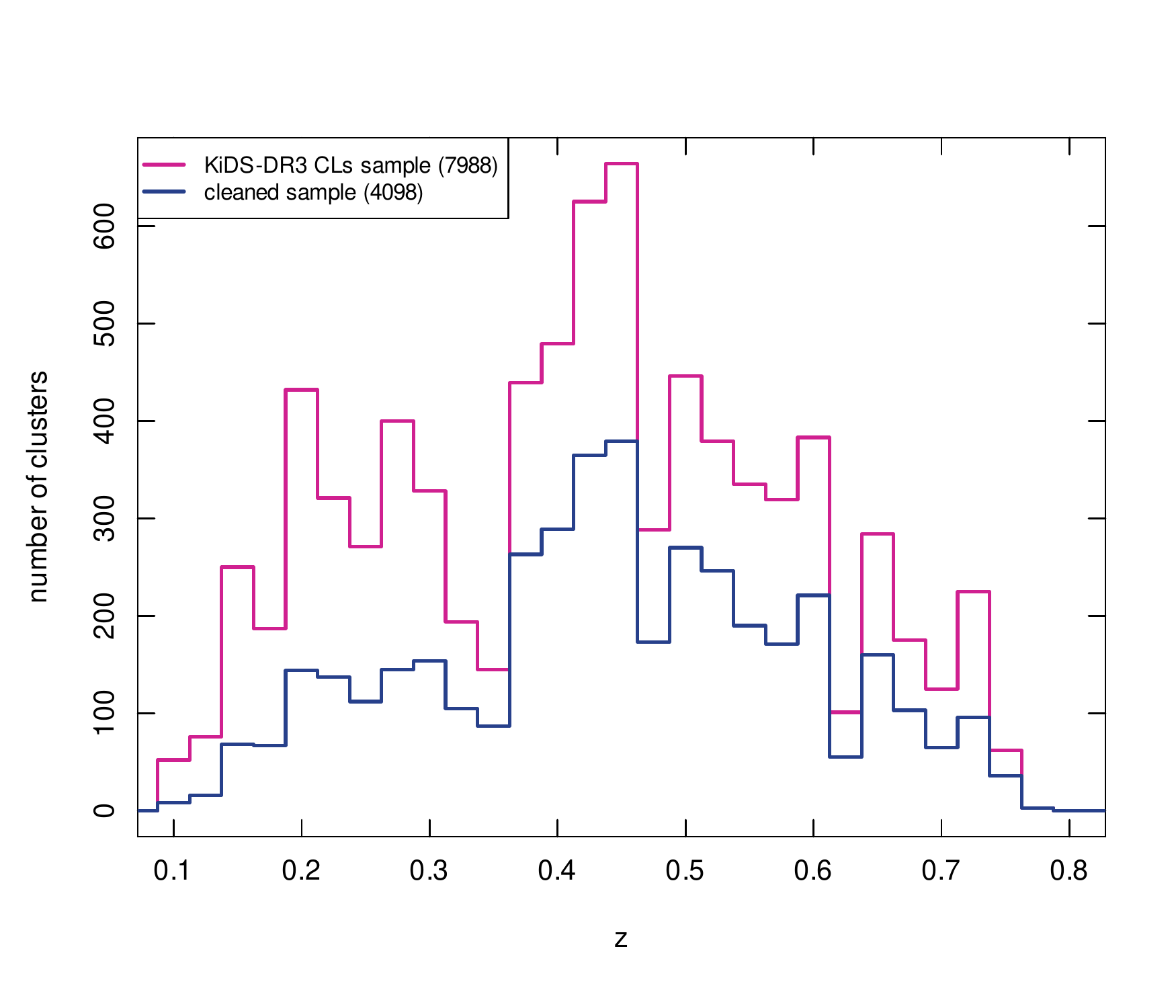}
     \caption{Redshift distribution for the KiDS-DR3 cluster sample (7988 clusters - in magenta) and for the sub-sample extracted for the purposes of this study (4098 clusters - in dark blue). } 
     \label{zdistr}
\end{figure}

\begin{figure*}
   \centering
   \includegraphics[width=8.cm]{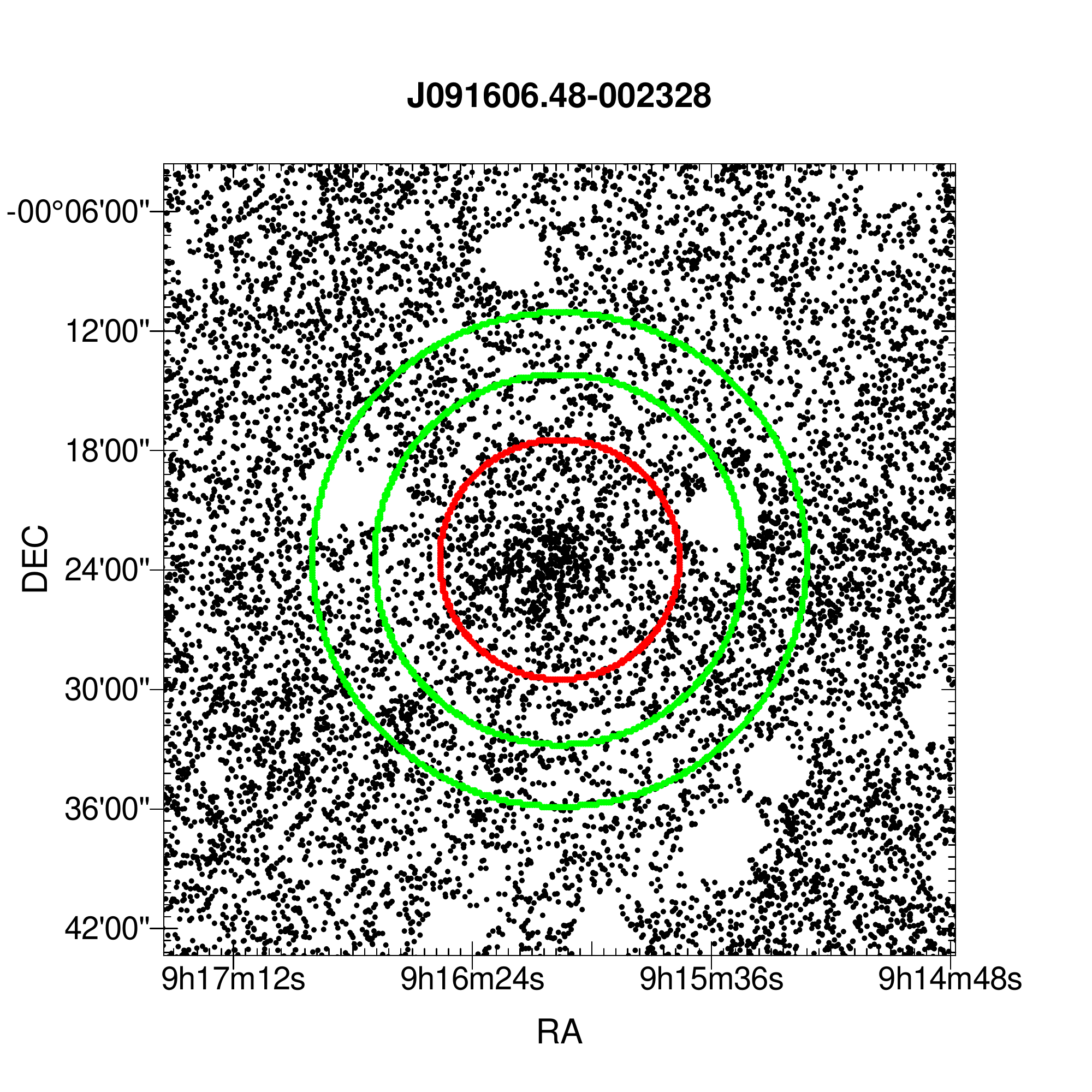}
   \includegraphics[width=10.cm]{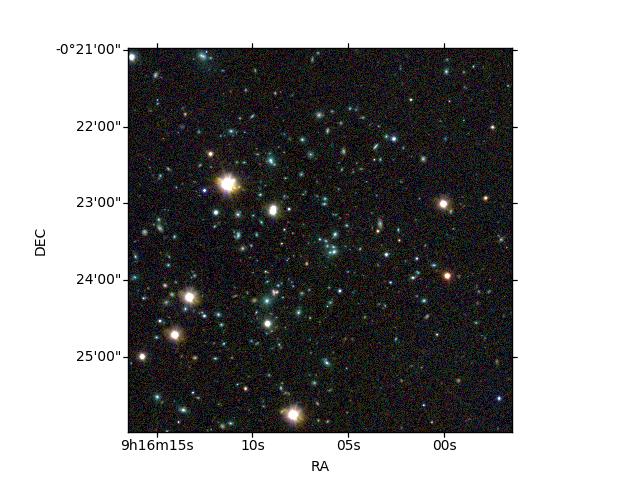}
     \caption{AMICO-KiDS detection AK3 J091606.48-002328 or Abell~0776. Left panel: Galaxy map of AK3 J091606.48-002328 (RA=139.027, DEC=$-$0.3913, S/N=9.68, z=0.37, $\lambda_{\star}$=137): the region spans an area of $\sim$~50$\arcmin \times$50$\arcmin$. 
     The black dots represent the galaxies brighter than $r_{lim}$=24.0. The red circle encloses the central area within $r_{200} \sim$~6$\arcmin$; and the two green circles delimit the area over which the background is estimated (in this case, an annulus with radii from 12$\arcmin$ to $\sim$~18$\arcmin$). 
     The white holes in the galaxies distribution indicate masked areas.
     Right panel: Colour composite $(g, r, i)$ image relative to AK3 J091606.48-002328. The stamp shows a $\sim$~6$\arcmin \times$6$\arcmin$ region centred on the location identified by AMICO.} 
     \label{Fig1}
\end{figure*}

When applied independently, the constraint on edges proximity eliminates 3622 candidates and the MASKFRAC constraint discards 3134 against the complete original catalogue of 7988 clusters. If we jointly consider the two constraints, 3890 clusters are dropped and we obtain a clean sample of 4098 clusters in the same redshift range  ($0.1<z<0.8$), as the parent range. 
In Fig.~\ref{zdistr} we show the redshift distributions of the clean sample  and of the whole KiDS-DR3 sample.

\begin{figure}
   \centering
   \includegraphics[width=9.cm]{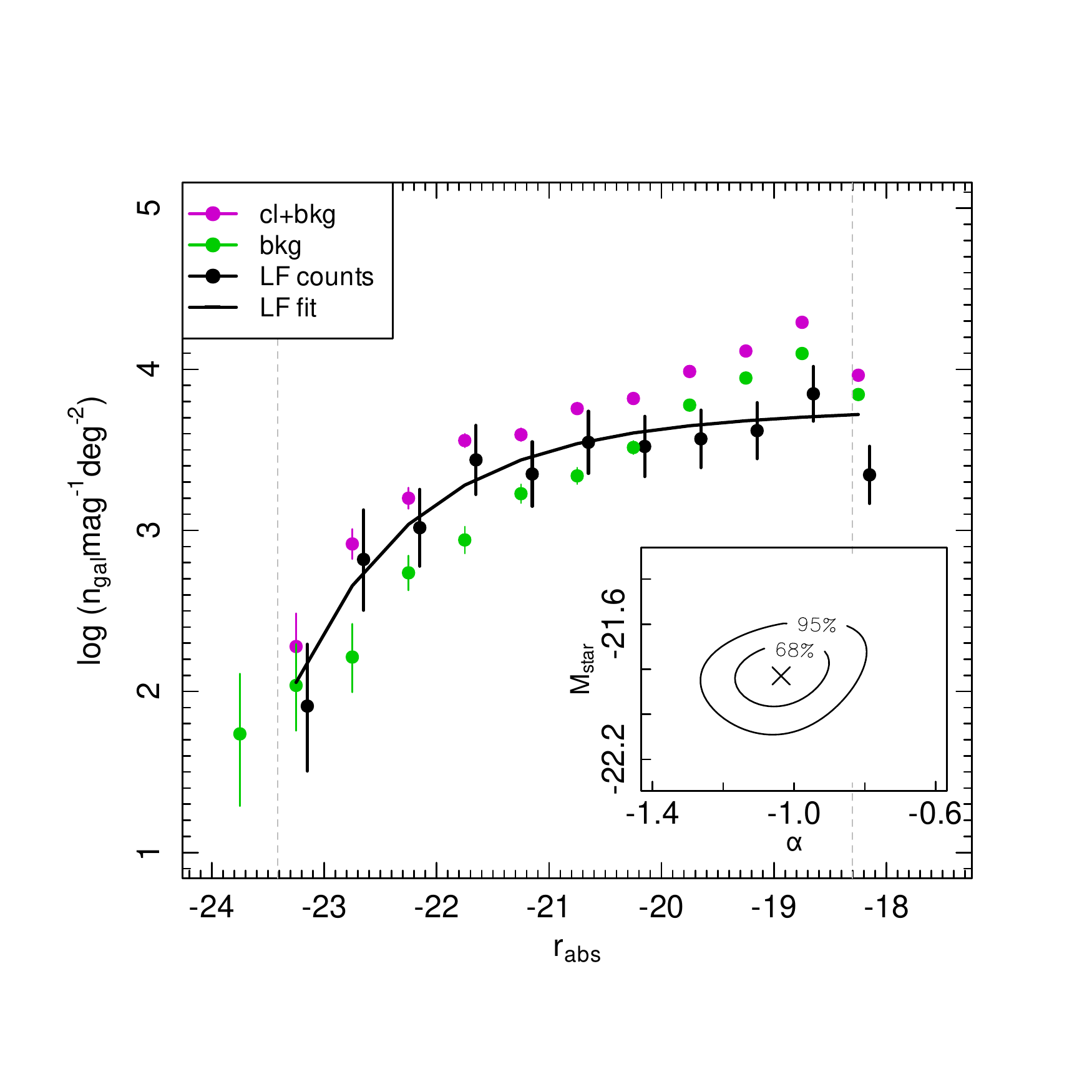}
     \caption{Luminosity function for AK3 J091606.48-002328. 
     The magenta dots represent the total counts coming from the region inside $r_{200}$; the green dots represent the background counts coming from the background annulus only; and the black dots refer to the background subtracted counts, fitted by a Schechter function (black solid line) in the magnitude range delimited by the two grey vertical dotted lines. Inset: The confidence plot for the values of $\alpha$ and $M_{\star}$ resulting from the fit; the contours at $1\sigma$ and $3\sigma$ are reported.} 
     \label{Fig2}
\end{figure}

\subsection{Single cluster LF}
\label{SingleLF}

To estimate the total counts (cluster+background) over the central region we chose 
$r_{in}$=1.2~$r_{200}$, accounting in this way for the uncertainty in the derivation of the cluster centre in a conservative way. The value 
$r_{200}$ is computed from the mass $M_{200}$ assuming a NFW cluster profile and $M_{200}$ is the cluster mass estimated from richness by means of the scaling relation calibrated in \citet{Bellag19}.

Before computing single cluster LF, for each cluster all the galaxies brighter than the BCG identified by \citet[][]{MarioBCG} and within $r_{in}$ were excluded. Moreover, we removed from the background annulus every over-density detected by AMICO falling into it.

As an example, in Fig.~\ref{Fig1} we show the density map (left panel) of the AMICO-KiDS detection AK3~J091606.48-002328 (Abell~0776),  detection with the highest signal-to-noise ratio (S$/$N=9.68) in our sample and an estimated mass of $M_{200}$=(6.2$\pm$3.8)$\times 10^{14}$M$_{\odot}/h$. 
We plot the spatial distribution of the galaxies (black dots) brighter than $r_{lim}$=24.0; we mark the cluster area (red circle) within $r_{in}\sim$~1.9 Mpc and the background annulus (green circles) that for this cluster has radii from 3.8 Mpc to 5.7 Mpc.
In the right panel, we show the colour composite $(g, r, i)$  stamp centred at the peak position identified by AMICO.

The single cluster LFs are computed by subtracting from the galaxy counts in magnitude bins of $0.5$~mag, the background counts normalized to the cluster area. 
In Fig.~\ref{Fig2}, we plot the magnitude distribution of all the galaxies belonging to the region inside $r_{200}$, the magnitude distribution of the background galaxies belonging to the annulus and the resulting cluster LF after the background subtraction, which is fitted by a Schechter function. The best-fit parameters $\alpha$ and $M_{\star}$ are $-$1.04$\pm$0.14 and $-$21.83$\pm$0.21, in agreement within the errors with the expected values ($\alpha$=$-$1.06 and $M_{\star}$=$-$21.79) at this cluster redshift by the \cite{Hennig} model.
The confidence regions at 1$\sigma$ and 3$\sigma$ for the values of $\alpha$ and $M_{\star}$ resulting from the fit, are reported in the inset in Fig.~\ref{Fig2}.

\subsection{Stacked LF}
\label{StackedLF}

Thanks to the sensitivity and the reliability of the AMICO code, our clusters sample spans a wide range of richness/mass, including groups of low mass, down to few $10^{13} M_{\odot}/h$.
Because of the low galaxy number counts in the low-mass or high-redshift groups, we cannot perform individual studies for these systems.
In the following, we perform a stacked analysis of the LF, in bins of cluster redshifts and richness. In this way, it is also possible to make a robust analysis of the LF for clusters with low richness or high redshifts, where the LF of individual clusters would be too noisy.

To build the stacked LF we transform the apparent magnitude in absolute magnitude according to 
\begin{equation}
    M_{abs}=m+5-5 log_{10}\left(\frac{D_L}{Mpc}\right)-kcorr,
\end{equation}
where $D_L$ is the luminosity distance and $kcorr$ the $k$ corrections, computed by {\it EzGal} (\cite{ezgal}), assuming that the main population of our clusters are passive evolving galaxies at the cluster redshift. Evolutive correction is not applied since we want to investigate this.

In this work, the stacking of the LF is done as proposed by \cite{Garilli99}, since the \cite{Colless89} method was shown in \cite{2018Ricci} to assign more weight to the poor clusters.  Following these prescriptions, the stacked LF is built by summing the cluster galaxies in absolute magnitude bins and scaling it by the richness of their parent clusters as follows: 
\begin{equation}
   N_{cj}=\frac{1}{m_{j}} \sum_{i} {N_{ij}}\cdot{w_{i}}, 
\end{equation}
where $N_{cj}$ is the number of galaxies in the $j$th absolute magnitude bin of the stacked LF;  $N_{ij}$ is the number of galaxies in the $j$th bin of the $i$th cluster LF; $m_j$ is the number of clusters with the limiting magnitude deeper than the magnitude of the $j$th bin; and $w_i$ is the weight of each cluster, given by the ratio of the number of galaxies of the $i$th cluster to the number of galaxies brighter than its magnitude limit in all clusters with fainter magnitude limits.
The formal error in the stacked LF is computed as 
\begin{equation}
  \delta N_{cj}=\frac{1}{m_{j}} \sqrt{\sum_{i} {N_{ij}}\cdot{w_{i}^{2}}} .
\end{equation}

\subsection{Colour selection}
\label{color}

To constrain the evolutionary scenarios of cluster galaxies, we studied the two main galaxy populations, red passive and blue star forming galaxies, separately. We applied the colour selection of \cite{MarioBCG} to derive the LFs of our cluster sample for the overall population as well as for the red and blue galaxies population. 
In \cite{AndreonBO} it is shown that identifying the red galaxies as those inside the colour stripe around the red sequence, because of colour evolution with redshift, introduces a bias so that at higher redshifts red galaxies could be wrongly classified as blue, spuriously enhancing the Butcher-Oemler effect \citep{B&O}. To circumvent this problem, \cite{MarioBCG} derived two models for galaxies E and Sa by using $EzGal$ \citep{ezgal}; on the colour-magnitude diagram the colours of the ellipticals E and the Sa at the cluster redshift indicate a strip where red galaxies lie. Instead, blue galaxies are those lying below the colour of the Sa.  To account for the quality of the red/blue galaxies separation, we change the colour at higher redshift: $g-r$ for clusters with $z <$0.4 and  $r-i$ for clusters with $z>$0.4.

\section{Results} \label{result}

In this section, we present our results. 
We first study the evolution of the stacked LFs with the redshift (Sect.~\ref{LF in redshift bins}) and the dependence on richness/mass (Sect.~\ref{LFinrichbin}). Then, in Sect.~\ref{LF2D}, we analyse the joint dependence on these two parameters in the 2D space of $z$ and $\lambda_{\star}$. With this procedure, we want to take into account the differences of mass distribution that occur at different redshifts. For example, at low redshift the low-mass clusters population dominates because rich clusters are rare, while at higher redshift high-mass clusters should be more detectable and, therefore, better sampled.

We divided the redshift range in four bins [0.1,0.32[, [0.32,0.46[, [0.46,0.52[, and [0.52,0.8[, as well as the $\lambda_{\star}$ range [0,15[, [15,28[, [28,50[, and [50,$\infty$[ (see Table \ref{table:1} and \ref{table:2}).  This partition was a compromise among the following criteria: bins in richness were chosen to enclose roughly the same number of clusters and bins in redshift were defined to be approximately equally spaced. 
The only exception are the bins on the highest redshift and richness values that contain fewer clusters, with a difference from several hundred to few tens. The lower number of clusters for these bins is compensated by the large richness, that is a large pool of galaxies, which guarantees solid statistical results. 
In the sample that we used for the LF estimate,  which encompasses about half of the AMICO-KiDS clusters, we only have two clusters of richness $\lambda_{\star}>100$ in the lower redshift bin (nine clusters in the whole DR3 catalogue). In terms of mass, there are 33 clusters with $M_{200} \geq 10^{14.5}$  (133 in the whole AMICO-KiDS-DR3 catalogue).

The Schechter fits of the stacked LFs for the total galaxies, red (ET) and blue (LT) are performed by the R package {\it nlm}. Errors are Poissonian and are computed using the formulae of \cite{Gehrels1986}.

\begin{figure*}
   \centering
   \includegraphics[width=18cm]{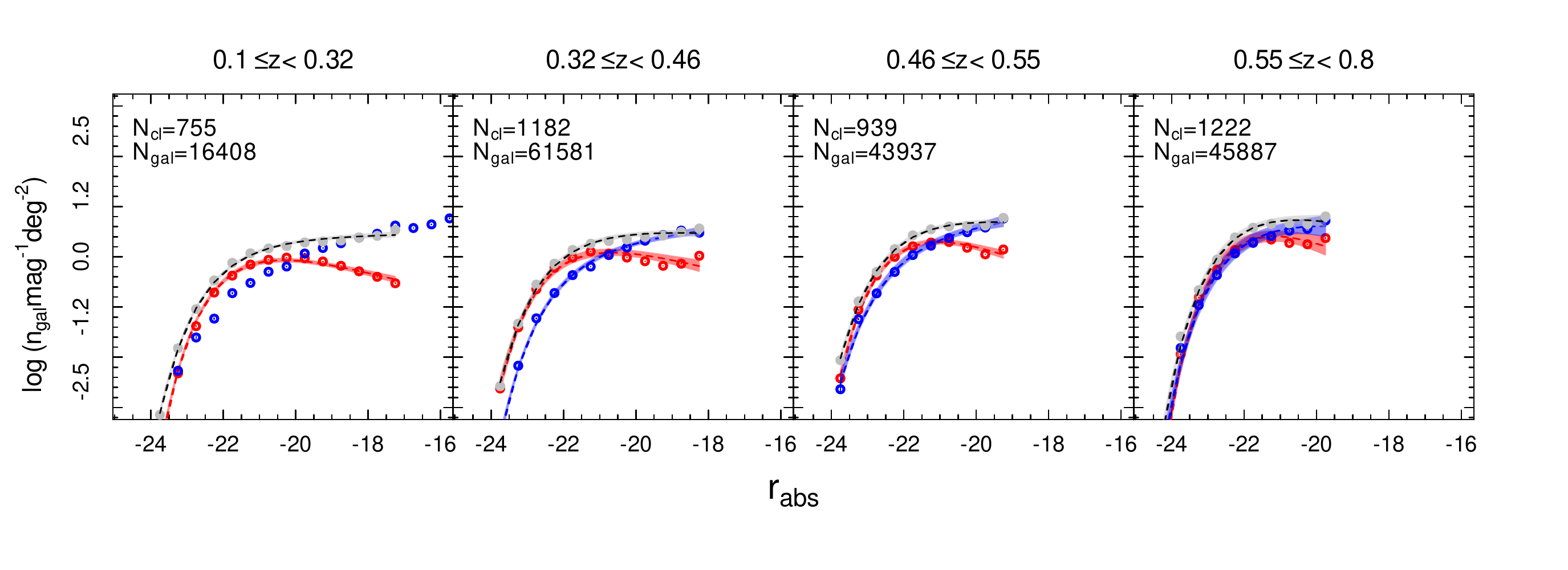}
   \caption{ Redshift evolution of the LFs for stacked  galaxy clusters in bins of $z$. 
   The grey, red, and blue dots correspond to the total, red, and blue galaxies LF counts normalized to $1$ deg$^2$, respectively. For each panel, the number of clusters contributing to the stacking ($N_{cl}$), the total number of galaxies inside $r_{200}$ belonging to these clusters ($N_{gal}$), and the redshift range are reported. The dashed lines (assuming the same colour code) are the fit to the Schechter function; the shaded regions enclose the $97.5\%$ confidence interval. The LF counts are plotted in the two cases of fits that do not converge (left panel, blue galaxy populations), but not the Schechter curves.}
    \label{LFz}
    \end{figure*}

\begin{figure}
   \centering
   \includegraphics[width=8cm]{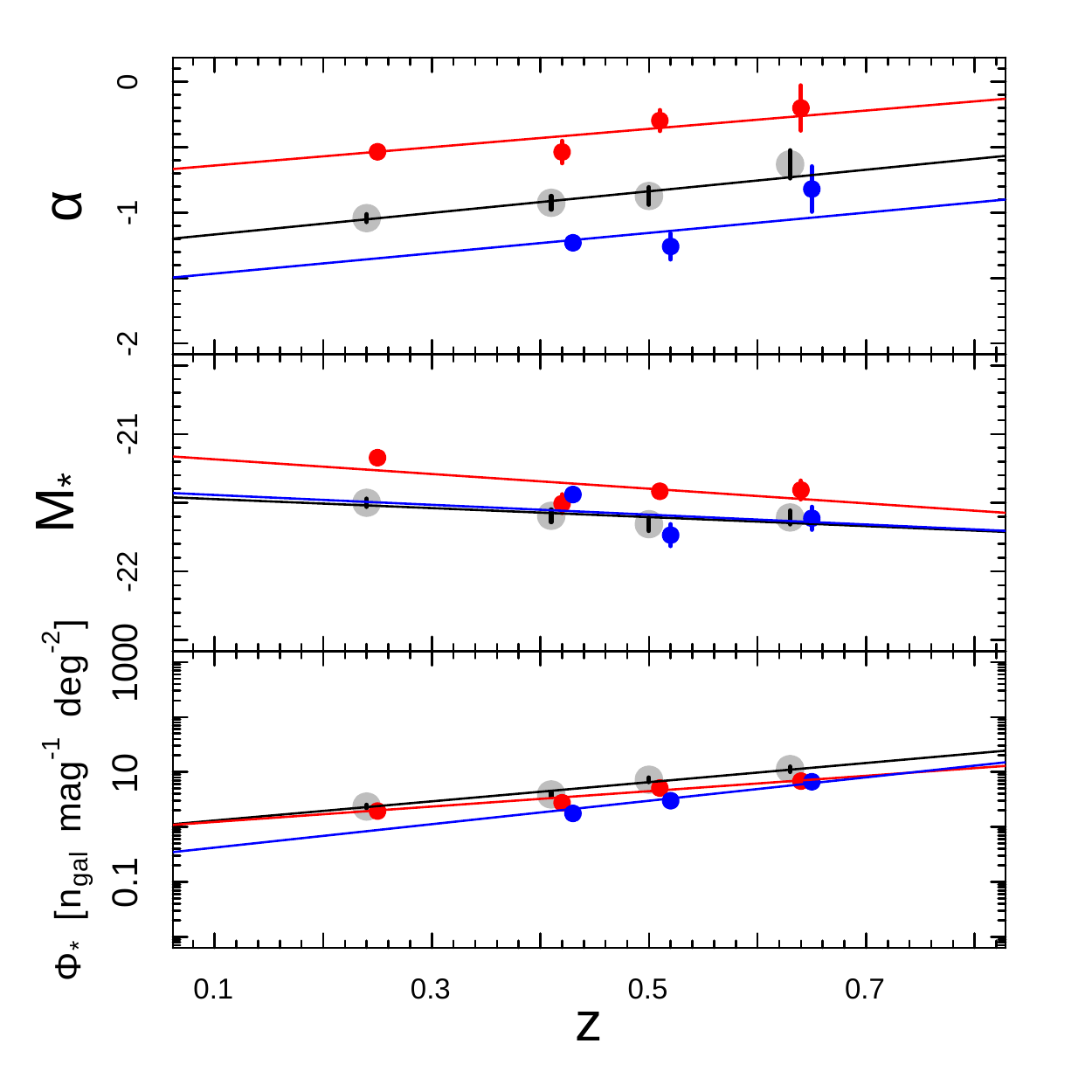}
   \caption{Redshift evolution of the Schechter fit parameters for all galaxies (grey dots), red galaxies (red dots), and blue galaxies (blue dots). The trend lines for all (black lines), red, and blue galaxies are overplotted.}
    \label{ztrend}
    \end{figure}
    
\begin{figure*}
   \centering
   \includegraphics[width=18cm]{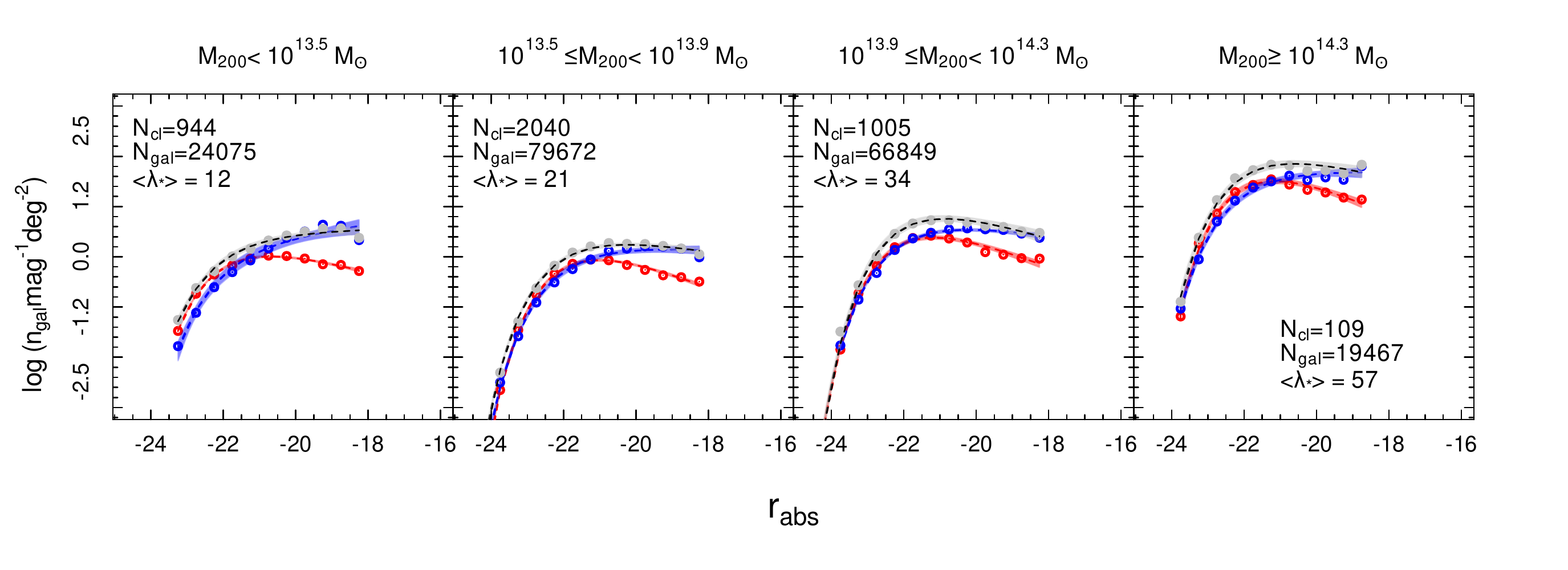}
   \caption{Richness/mass dependence of the stacked LFs. 
   The legend is the same as in Fig.~\ref{LFz}.
   Moreover, we report the median $\lambda_{\star}$ in the bin and the corresponding mass range.}
    \label{LFrichness}
    \end{figure*}
    
\begin{figure}
   \centering
   \includegraphics[width=8cm]{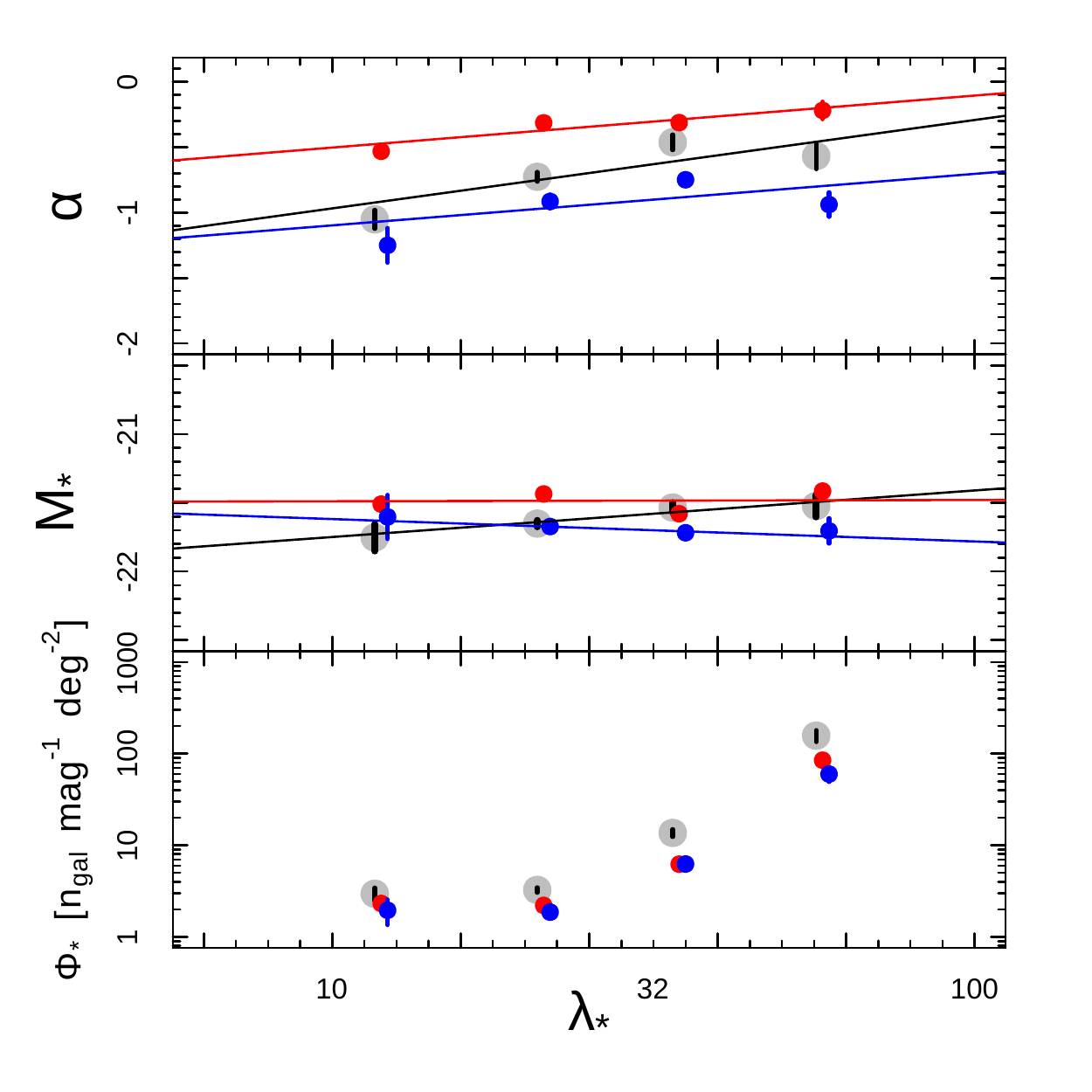}
   \caption{Dependence of the Schechter fit parameters on richness/mass for all galaxies (grey dots), red galaxies (red dots), and blue galaxies (blue dots). The trend lines for all (black lines), red, and blue galaxies are overplotted, except for the lower panel where $\Phi_{\star}$ is increasing but with a clear non-linear trend.}
    \label{richnesstrend}
    \end{figure}

\subsection{Luminosity function redshift evolution}
\label{LF in redshift bins}

In this section, we analyse the LF dependence on redshift. 
The redshift evolution of the LF is presented in  Fig.~\ref{LFz}.  The points represent the counts: grey dots for the total counts, red for ET and blue for LT counts; the shaded coloured regions show the 97.5$\%$ confidence intervals around the Schechter fitted curves. 
The confidence regions are obtained by bootstrap resampling with 1000 iterations.  The Schechter fit parameters are listed in Table~\ref{table:1}.
In some cases (see the blue galaxies in the bin with the lowest redshift and mass), the fit did not converge to a single Schechter function \citep[see][]{Popesso05}, and we lack the values of the parameters $\alpha$, $M_{\star}$, and $\Phi_{\star}$.

If we look at the brighter part of the LF, $-24>r_{abs}>-20$, we note that the red counts are above the blue counts and intersect at increasingly bright magnitudes with redshift. The area bounded by red and blue LFs, which is related to the ratio of  bright red to bright blue galaxies, shrinks down to disappear at higher redshifts. In parallel, the blue faint-end slope, which is steeper than the red one at low redshift, approaches the red faint-end slope for increasing $z$, becoming shallower. This could be interpreted as the decline of the faint blue fraction, which at higher redshifts approaches the faint red galaxy fraction. 

In Fig.~\ref{ztrend}, we plot the trends with redshift of $\alpha$, $M_{\star,}$ and $\Phi_{\star}$ for the red, blue, and total populations; one $z$ bin corresponds to each point. To quantify the significance of these trends, we performed a linear fit represented in the figure by the overplotted red, blue, and black lines.
The  coefficients and the significance levels  are listed in Table \ref{table:sigz}.
We note that, in the upper panel showing the $\alpha$ trend with $z$, the red dots are above the blue and grey dots: this means that the red galaxies have a distinctly less steep faint-end slope than the blue and total populations across the whole redshift range. The value $\alpha$ increases with redshift for the red, blue, and total populations with a low significance ($\lesssim$~2~$\sigma$, see Table \ref{table:sigz}), the LF slope becoming less steep; it flattens for the total and blue population and becomes shallower for the red galaxies.   
In the middle panel, we show that the absolute magnitude characteristic parameter $M_{\star}$ decreases for the red galaxies ($<$~1.2~$\sigma$) and remains constant for the other two populations. 
The normalization $\Phi_{\star}$ mildly increases with redshift for the total and red populations at $\sim$~3~$\sigma$ and for the blue population at 1.4~$\sigma$.  
The density of the red galaxies, which are larger (the red dots above the blue dots) at low to intermediate redshifts, approaches the blue one at high redshift.

\subsection{Luminosity function dependence on richness}
\label{LFinrichbin}

In this section, we analyse the  LF dependence on richness. 
In this case, the clusters in each richness/mass bin and at all redshifts are stacked together. 
The Schechter fit parameters are listed in Table \ref{table:2}.
In the plots of Fig.~\ref{LFrichness}, We see very small difference in the area bounded by the red and blue LFs at bright magnitudes in bins of increasing richness/mass; the blue faint-end slope is steeper than the red one at low richness and approaches it as richness increases.  

In Fig.~\ref{richnesstrend}, we plot the trends as a function of richness of $\alpha$, $M_{\star}$ and $\Phi_{\star}$, for the red, blue, and total populations; one $\lambda_{\star}$ bin corresponds to each point. The  red, blue, and black lines represent the linear fit, whose coefficients and significance are listed in Table \ref{table:sigz}. 

As for the redshift evolution, the red LF is less steep than the total/blue LF (the red dots are above the blue dots and the grey dots are in the middle) independent of richness. 
The parameter $\alpha$ increases (the LF slope becomes less steep) with richness/mass with a significance level weaker for blue/red galaxies ($\sim$~1~$\sigma$), and a significance of 1.7~$\sigma$ for the total population. 
Moreover, at low richness the total population follows the red population in the bright part and the blue population  
in the faint part; in the left panel of Fig.~\ref{LFrichness}, the grey and  blue dots are very close; as the richness increases, the total faint-end slope approaches the red one (in the last panel of Fig.~\ref{LFrichness}).  
The characteristic absolute magnitude parameter $M_{\star}$ is compatible with no richness/mass dependence for the red, blue, and total populations. 
The characteristic density parameter $\Phi_{\star}$ depends on richness/mass for the total galaxy population, as expected since it is directly related to richness. 

\subsection{Luminosity function in richness/redshift 2D bins: Breaking the degeneracy between redshift and mass}
\label{LF2D}
Our sample comprises a large number of clusters over a wide range of mass and redshift, which allows us to disentangle possible effects of the blending of different richnesses and redshifts dependences. 
To do this, we split each redshift bin into four bins of $\lambda_{\star}$ already listed and used in Sect.~\ref{LFinrichbin}, and we built a set of 16 2D bins of both redshift and richness, in which we derive the stacked LFs and the Schechter parameters $\alpha$, $M_{\star}$, and $\Phi_{\star}$. 

\begin{figure*}
  \centering
   \includegraphics[width=18cm]{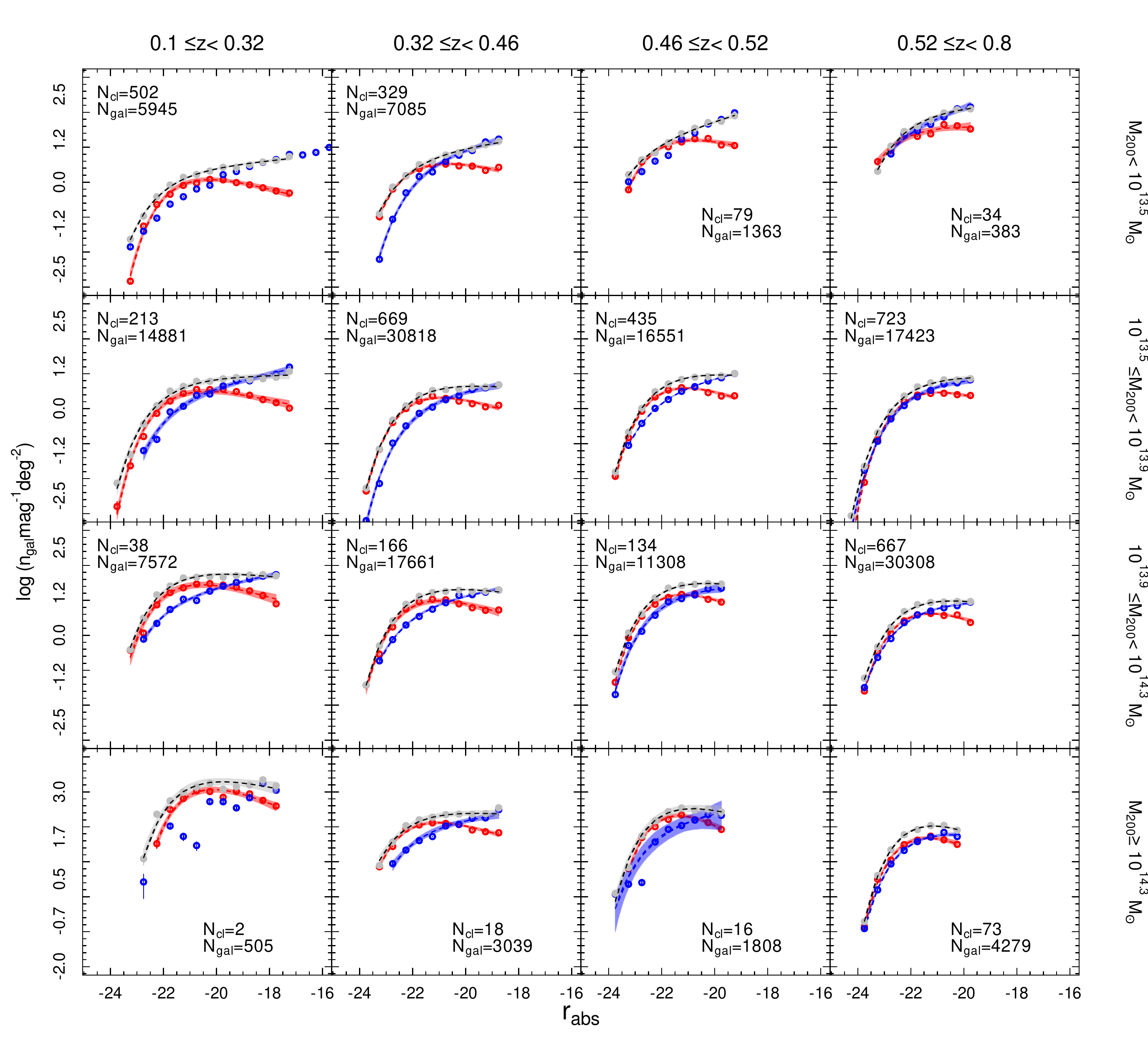}
   \caption{Redshift and richness/mass evolution of the stacked LFs in bin of $z$ and $\lambda_{\star}$ for the total population of galaxies (black dots), red galaxies (red dots), and blue galaxies (blue dots); LF counts are normalized to 1 deg$^2$. The dashed lines are the fit to the Schechter function; the shaded regions enclose the $97.5\%$ confidence interval. In those few cases where the LF cannot be fitted to a Schechter, at least the LF counts are plotted. }
   \label{LFzrich}
   \end{figure*}

\begin{figure*}
   \centering
  \includegraphics[width=16cm]{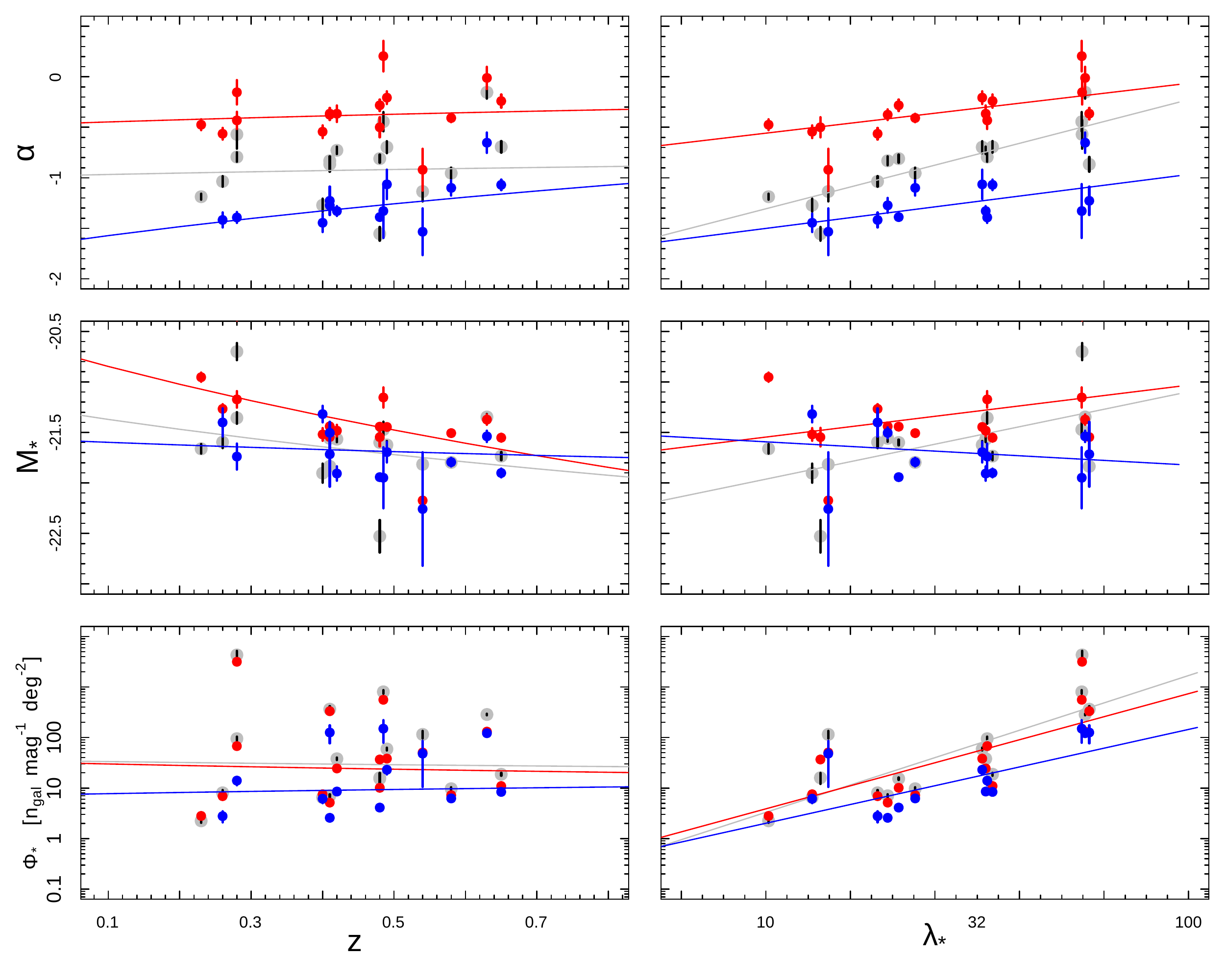}
  \caption{Evolution of the Schechter fit parameters with redshift and richness/mass for all galaxies (grey), ETs (red), and LTs (blue). The trend lines represent the eq. (\ref{eq10ofRicci}) at fixed values of $z$=0.35 for the richness dependence (right panels) and $\lambda_{\star}$=22.5 for the redshift dependence (left panels).}
   \label{ztrend-2dbins}
   \end{figure*}

The stacked LF counts, normalized to 1~deg$^2$ area and corresponding to a given 2D bin, are shown in each panel of Fig.~\ref{LFzrich}. From left to right, each column represents bins of increasing redshift, whereas each row from top to bottom corresponds to a given range of growing richness/mass.
The stacked LFs are plotted for the total population of galaxies, ET, and LT galaxies. As in the previous sections, we overplot the fits to the Schechter function and the shaded regions enclosing the 97.5$\%$ confidence interval. 
The fit parameters are listed in Table \ref{table:3} and plotted in Fig.~\ref{ztrend-2dbins}  as a function of the median redshift (left panels) and richness (right panels) for the total, red, and blue galaxy populations with the usual colour code. The left column of panels, illustrating the redshift evolution, is analogue to Fig.~\ref{ztrend}, where this time each data point has been split in four bins of $\lambda_{\star}$; whereas the right column, showing the Schechter parameters dependence on our mass proxy, $\lambda_{\star}$, is akin to Fig.~\ref{richnesstrend}, whose data points were also split into four $z$ bins. 

We then fit the evolution with redshift and the richness dependence of the Schechter parameters  following the prescriptions of \citet{2018Ricci}, who assumed a linear dependence from the two independent variables and fit them conjointly. With this analysis we disentangle the interplay between richness and redshift. Following \citet{2018Ricci}, we consider the relation
\begin{equation}
\label{eq10ofRicci}
Y = a \cdot log(1+\langle z \rangle) + b \cdot log(\langle\lambda_\star\rangle) + c   
,\end{equation}
where $Y$ takes on the meaning of each of the three parameters of the Schechter function, $\alpha$, $M_{\star}$ e $\Phi_{\star}$; $\langle z\rangle$ and $\langle \lambda_{\star}\rangle$ are the median redshift and median richness, respectively, of the given bin; $a$, $b$, and $c$ are the coefficients of the linear model to be constrained. 
We fit our sample of 16 data points, corresponding to the redshift-richness 2D bins, adopting equation \ref{eq10ofRicci} by using the R package "lira", which performs Bayesian linear regression \citep{MSBayes}. The results are listed in Table \ref{table:sigzl}.
In Fig.~\ref{ztrend-2dbins}, we overplot the best-fit lines graphically representing equation \ref{eq10ofRicci} 
at fixed values  of $z$=0.35 for the richness dependence (right panels) and $\lambda_{\star}$=22.5 for the redshift dependence, respectively (left panels).

The parameter $\alpha$ still increases with redshift for the blue population at 2~$\sigma$, while for the red and total populations the redshift dependence seems to be negligible when the $\lambda_{\star}$ variable is introduced. 
The absolute magnitude characteristic parameter $M_{\star}$ decreases for the red and total populations (at 2.1~$\sigma$ and 1.5~$\sigma$) and remains constant for the blue galaxies, a trend compatible with passive evolution for the red and total populations and with no evolution for the blue galaxies. 
The splitting in richness bins makes the mild $\Phi_{\star}$ increase with $z$ we see in Fig.~\ref{ztrend} disappear. 
We would like to point out that the distance between red dots (always above the blue dots) and blue dots decreases with redshift. This means that the red density approaches the blue density as the redshift increases; a similar trend is found for the red/blue fractions of galaxies, which we discuss in Sect.~\ref{redandbluefrac}.

There is clear evidence of an $\alpha$ increasing (LF faint-end slope becoming less steep) with richness/mass for the total populations, significant at 5.2~$\sigma$, and for a less significant dependence for the red/blue galaxies. 
The parameter $\Phi_{\star}$ obviously increases with richness in all the three cases, whereas $M_{\star}$ becomes slightly fainter with richness/mass for the red galaxies and for the total population at a low level of significance ($\sim~1\sigma$), and does not depend on richness for the blue population.

\begin{figure}
   \centering
  \includegraphics[width=8.5cm]{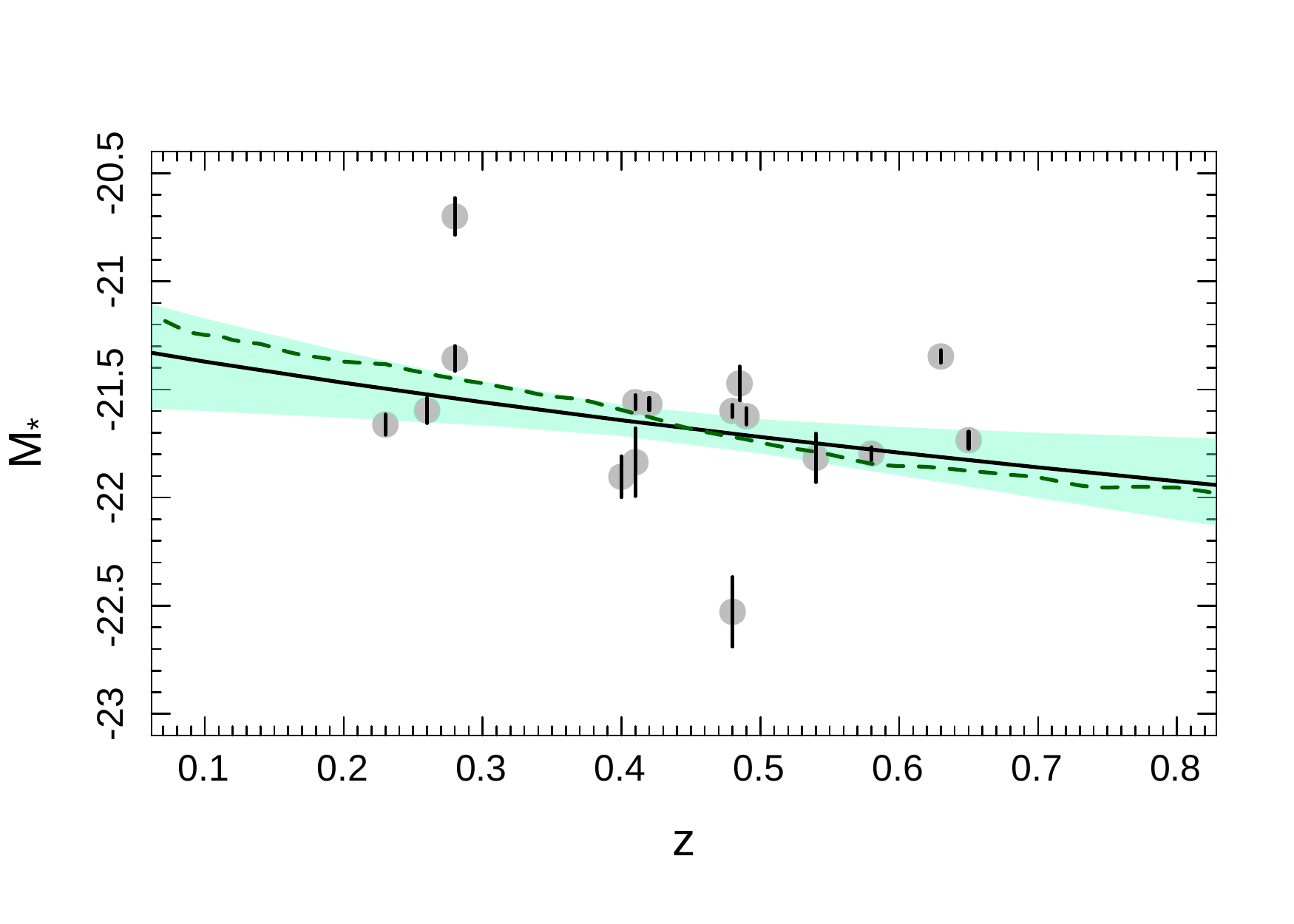}
  \caption{Comparison between the evolution model for $M_{\star}$ \citep{Hennig} (green dashed line), used as a reference in the detection algorithm, and the observed $M_{\star}$ evolution (grey dots). The 1$\sigma$ confidence region (shaded area) of the best fit (black line drawn for $\lambda_{\star}$=22.5) to our data is reported.}
   \label{mstar_evolution}
   \end{figure}

In Fig. \ref{mstar_evolution}, we show the comparison between the evolution model of $m_{\star}(z)$ \citep{Hennig}, adopted by the detection algorithm AMICO \citep{Bellag18}, and the observed $M_{\star}$ evolution. The 1$\sigma$ confidence region of the best fit  to our data is reported. Our estimate of the characteristic magnitude is compatible with \citep{Hennig}, that is a very mild redshift dependence.

 \subsection{Red and blue fractions}
 \label{redandbluefrac}
 
 The red/blue galaxy fraction is computed considering only the galaxies brighter than  $M_V = -20.5$ to sample the same region of the luminosity distribution for all the clusters. The reference magnitude $M_V = -20.5$ is converted to the observed apparent $r$--band magnitude by means of {\it EzGal}, assuming (as in Sect.~\ref{StackedLF}) that the main population of our clusters are passive evolving galaxies at the cluster redshift  \citep[see][for details]{MarioBCG}.
 Applying the statistical subtraction of the background by the central region,  for each cluster we obtain the galaxy counts in magnitude bins per unit area, we sum them over the magnitude range, and we multiply for the circular area encompassed by $r_{200}$ to obtain the total number of galaxies inside $r_{200}$ ($N_{gal,200}$). We perform this computation for the total sample and  for the red and blue galaxy samples, obtaining the number of all, red, and blue galaxies inside $r_{200}$ with luminosity up to $M_V = -20.5$. We define the red/blue fractions as the ratio of the red/blue galaxies to the total galaxies.
We note that these fractions are derived using statistical subtraction and this means that the number of red plus blue galaxies is not exactly equal to the total number of galaxies if computed in the way we describe above. This generally prevents the sum of the red and blue fractions from being equal to 1. 
We checked the difference between  the number of red plus blue galaxies and the total number of galaxies for each cluster in our sample and we find it to be on average of 1.5$\%$ and at maximum of 25$\%$ with only about 100 clusters exceeding 5$\%$. 

\begin{figure*}
  \centering
   \includegraphics[width=18cm]{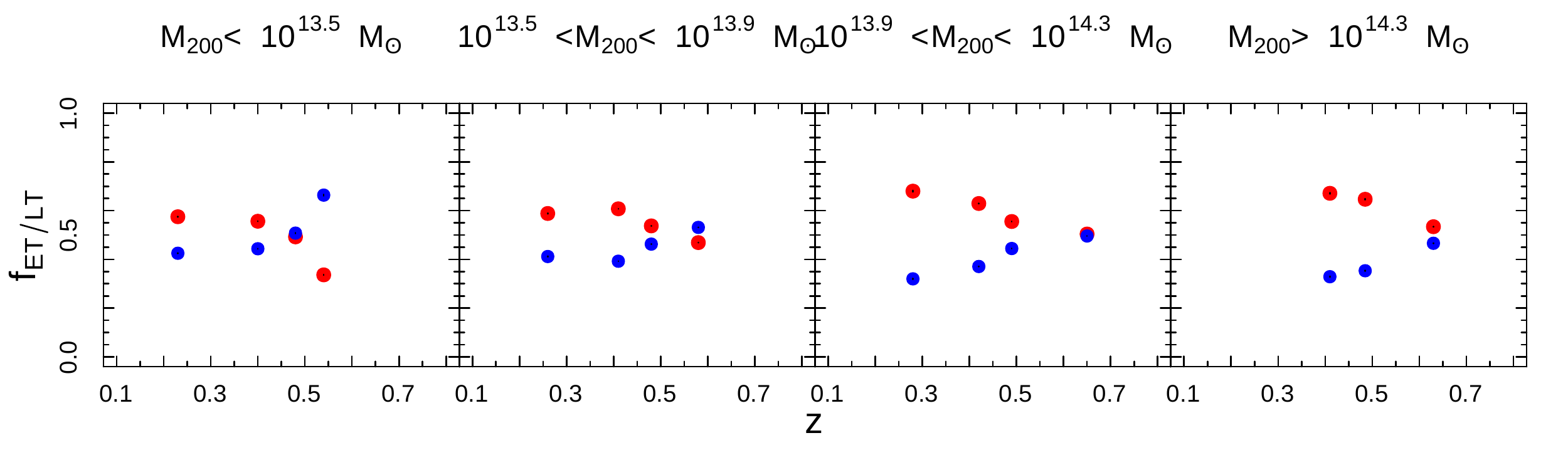}
   \includegraphics[width=18cm]{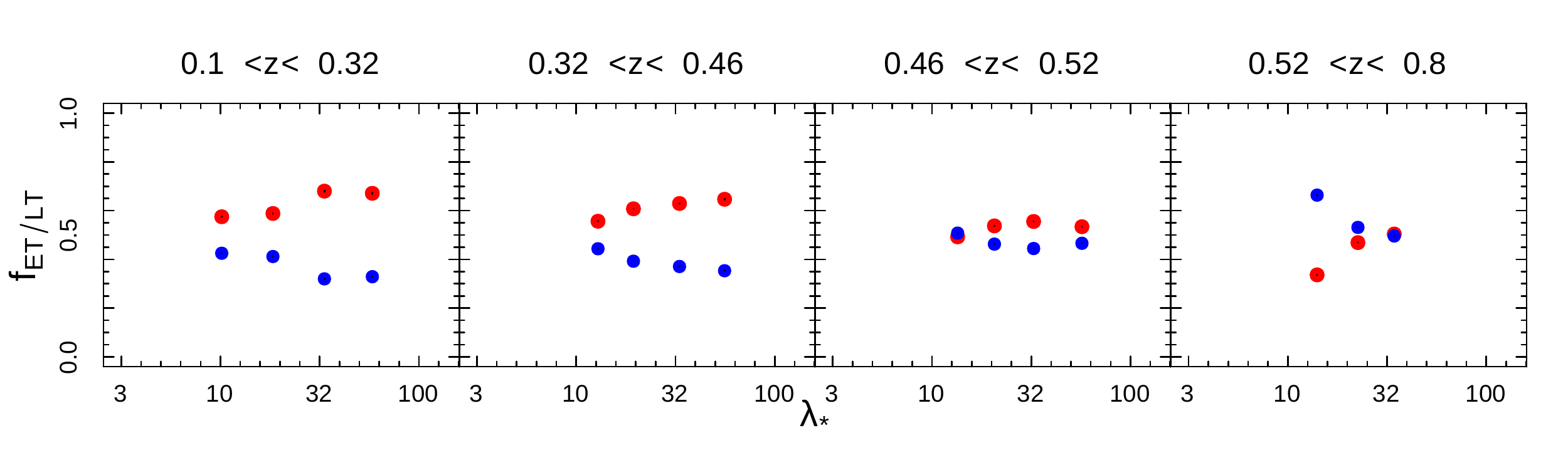}
   \caption{Red/blue galaxy fraction redshift evolution in richness/mass bins (upper panels) and red/blue galaxy fraction dependence on mass in redshift bins (lower panels). The red/blue dots represent the red/blue population, respectively. The error bars are smaller than the symbols size.  } 
   \label{RBfrac}
   \end{figure*}

In Fig.~\ref{RBfrac} we show the stacked ET/LT fractions in the 16 bins of ($z$,$\lambda_{\star}$) already used across this paper (see Table \ref{table:3}). In the upper panel, we represent the ET/LT fractions as a function of redshift in the four bins of richness/mass and in the lower panel, conversely, we plot the ET/LT fractions as a function of $\lambda_{\star}$ in the four bins of redshift.
We compare our results with \citet{MarioBCG}. These authors derived red/blue fractions from the same catalogue, but used member probabilities instead of  statistical subtraction, which we used in this work. 
Neverthless, the trends for our fractions are in agreement with \citet{MarioBCG}, who also find the red fraction to be higher than the blue fraction at low redshift and to decrease with $z$.  Similar results were also found by \cite{wen2018}, \cite{Sarron}, and \cite{Hennig}. 
The crossing point occurs at increasing redshifts in bins of increasing richness/mass.
The dependence on richness is in agreement with \cite{MarioBCG} in the different redshift bins except for high $z$ clusters: the red fraction is always higher than the blue fraction, but the mean distance between them decreases with $z$. For the higher redshift bin, the blue fraction is larger than the red fraction, but in this bin the results are more uncertain because of the larger errors in the colour classification of the galaxies and the small number of massive clusters (the dot corresponding to high richness is missing).

\section{Comparison with other studies}
\label{coparison_lit}

\begin{figure}
   \centering
   \includegraphics[width=8cm]{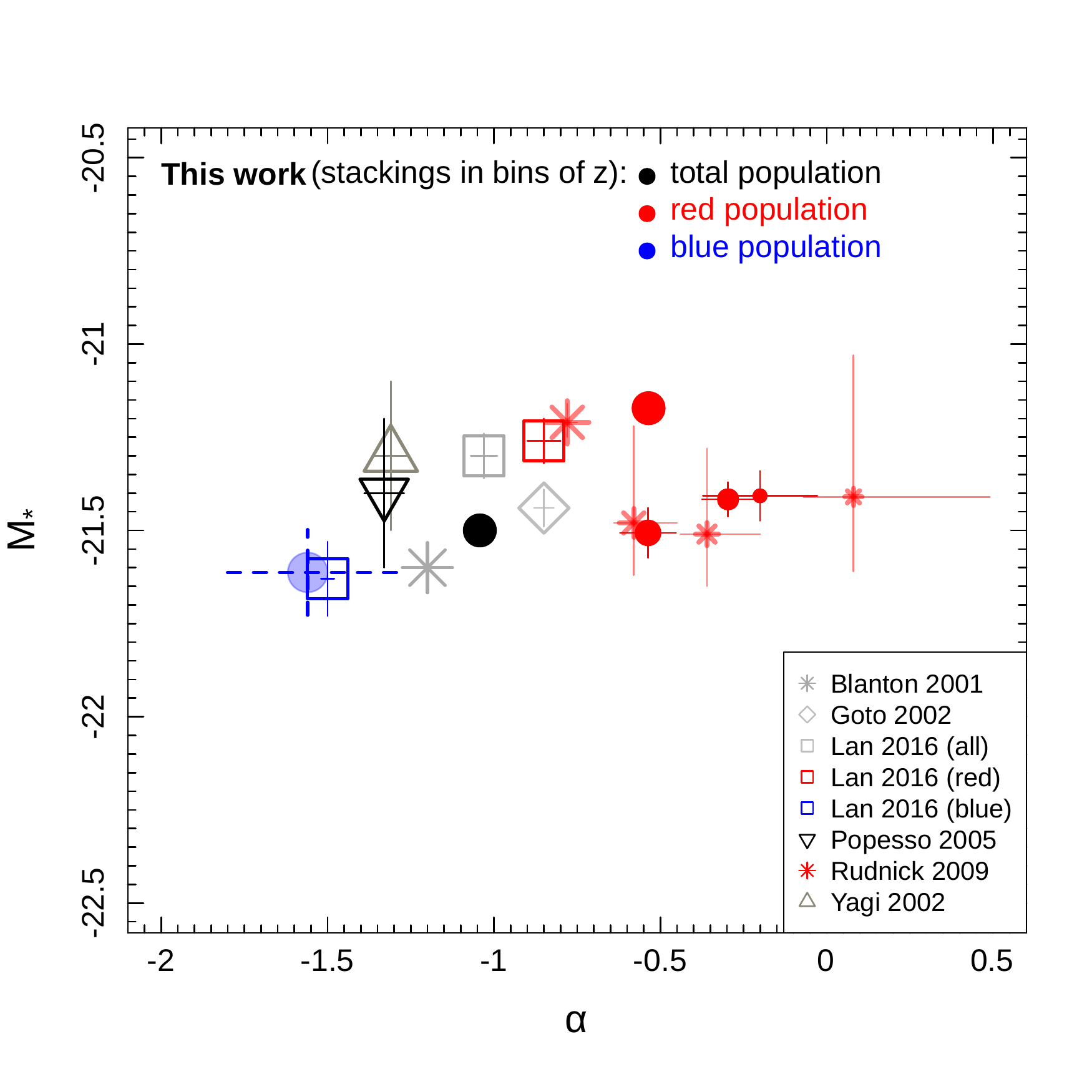}
   \caption{Comparison of our (filled symbols) derivation of the Schechter parameters with previous studies of composite LFs in the literature (empty symbols). Our red and total population parameters are those represented in Fig.~\ref{ztrend} and were obtained by stacking LFs in bins of $z$ at all richnesses; the single pale blue dot represents the estimate of our blue parameters by Eq.~\ref{eq10ofRicci}.
   The dimension of each symbol is inversely proportional to the median redshift of the $z$-bin.
   This also holds for the parameters from literature, where the used redshift is the median computed over the survey range.}
    \label{compareXXLS}
    \end{figure}

A proper comparison of our results with other studies is hampered by the differences in the survey properties (area and depth) and in the detection algorithms.  Most of these studies (excluding some very recent) have worked with much smaller samples than ours and have provided less reliable constraints, especially as concerns redshift evolution.

In Fig.\ref{compareXXLS}, in the plane $(\alpha,M_{\star})$ we compare our Schechter parameters (filled symbols) with other derivations from literature (empty symbols). As we adopt a cluster aperture of $r_{200}$ for our LFs, to make a reliable comparison we refer to those studies that map not only the central area of the cluster but a more extended region that is comparable with ours.
Our $M_{\star}$ (red circles, representing the red population stacked in bins of $z$ at all the  richnesses) agree with \citet{Rudnick2009}, who stack red-sequence galaxies in three bins in the range 0.4~$<z<$~0.8, plus another low-redshift bin coming from Sloan data, all for the Sloan $r$ magnitudes. The symbols size is inversely proportional to the median redshift of the $z$-bin.
Our low redshift $M_{\star}$ for the red population is also compatible with that of the passively evolving population derived by \citet{2016Lan} from SDSS data.
The comparison with the star-forming population of \citet{2016Lan} is  
unfortunately impossible because we do not have a low $z$ $M_{\star}$ for our blue population (see Sect.~\ref{LFz}); then, we plot the values of $\alpha$ and $M_{\star}$ estimated by the Eq.~\ref{eq10ofRicci} for $z=$0.03 (the average redshift of the \citet{2016Lan} sample) and over $\lambda_{\star} \in [10,100]$ (represented by the dashed error bars).  
In agreement with our results, \citet{2016Lan} find that $M_{\star}$ of the blue star-forming galaxies is brighter than that of the red passive galaxies.
Our parameters for the total population are compatible within the errors with other colour-independent LF derivations in clusters \citep{Goto2002,Popesso2005,Yagi2002} and also in the field \citep{Blanton2001} for low-redshift samples. 

As concerns the redshift evolution, we agree with \citet{Sarron}. 
These authors studied the evolution with mass and redshift of 1371 cluster candidates extracted from 154 deg$^2$ of the CFHT Legacy Survey using of the AMASCFI Cluster FInder. 
The sample has masses M$_{200} >$10$^{14}$M$_{\odot}$ and redshift $z \leq$ 0.7; they classify the galaxies in ET and LT by means of colours and derive the corresponding $i$-band LFs.
In the two mass bins we have in common, which are 10$^{13.9}$M$_{\odot} <$M$_{200} <$10$^{14.3}$M$_{\odot}$ and 10$^{14.3}$M$_{\odot} <$M$_{200} <$10$^{14.6}$M$_{\odot}$, there is consistency for the $\alpha$ and $M_{\star}$ trends of red, blue, and total populations. The main difference is that their ET $M_{\star}$ is brighter than the LT one of about half magnitude, whereas we find the opposite. This may reflect the fact that 
they sample a cluster region that is further inward than ours, in which there are not enough blue bright galaxies; we note that \citet{2016Lan}, who work on a more extended region around the cluster centre, found like us blue $M_{\star}$ brighter than red $M_{\star}$.

The $\Phi_{\star}$ trends are also compatible with \citet{Sarron}; at low $z$ the density of the red galaxies are higher than the blue galaxies and the red fraction approaches unity, whereas at higher redshifts the reds reduce and the blues increase, approaching each other. As concerns the mass dependence, these authors estimate an $\alpha$ dependence on mass (which we find) to be confirmed with a more suitable data sample and find (in agreement with us) no trend of the characteristic magnitude with mass.

Our findings are consistent with \citet{2018Ricci}. 
They investigate the LF of a sample of 142 X-ray selected clusters spanning a redshift range of about 0<$z$<1 and a mass range of $10^{13}$M$_{\odot}$<$M_{500}$<5$\cdot 10^{14}$M$_{\odot}$. On the Canada France Hawaii Telescope Legacy Survey (CHTLS) photometric galaxy catalogue, associated with photometric redshifts, these authors study the evolution of the galaxy luminosity distributions with redshift and richness, separately analysing the brightest cluster galaxy and non-BCG members.
They do not separate their sample into red and blue populations and therefore the comparison with our study concerns only the global population of galaxies.
As in our case, their results are compatible with no redshift evolution (except passive evolution) for all the Schechter parameters, and no richness dependence for the characteristic magnitude.
Moreover, their values for $\Phi_{\star}$ and $\alpha$ increase with richness in agreement with our results; for $\alpha$ it means that the slope becomes shallower with richness.

Our results are also in agreement with \citet{Zhang}, who derived the red galaxies LF of $\sim$100 X-ray selected clusters from the DES Science Verification data. The ranges of redshift and mass were 0.1<$z$<1.05 and 13.5$\leq$log$_{10}$(M$_{200}$)$\sim$<15.0, respectively.
These authors used a hierarchical Bayesian model to fit the cluster galaxy LFs with a Schechter function and find weak and statistically low significant ($\sim$~1.9~$\sigma$) evolution in the faint-end slope versus redshift and no dependence in $\alpha$ or m$_{\star}$ with their X-ray inferred cluster masses. 

Thanks to a mode-filtering technique for removing spatially variable backgrounds, \citet{Connor2017} produced very deep catalogues for the 25 CLASH \citep{CLASH2012} massive galaxy clusters at redshifts 0.2$\lesssim z\lesssim$ 0.9 obtained with the Hubble Space Telescope (HST). These authors measured the LF to depth of $\sim$5 mag fainter than $M_{\star}$ for the entire sample and to $\sim {M}_{\star}$+7 for the lowest-redshift clusters. 
They found a passively evolving value of $M_{\star}$ with redshift and no significant evolution in the population of faint galaxies, consistent with our work.

In summary, our results agree with several authors who find no evolution in the faint-end slope of red-sequence LFs with redshift  \citep[e.g.][]{Andreon2006,dePropris2007,dePropris2013,Lin2006,Muzzin2007,Mancone2012,Chiu2016} in our (or higher) redshift range. 

The already cited work of \citet{2016Lan} analysed the LF dependence on mass. These authors find no trend or steepening of $\alpha$ with mass, as they fit, respectively, the bright ($M_r<$~$-$18) and the faint ($-$18~$<M_r<$~$-$12) LF part with a double Schechter and this conflicts with our finding of $\alpha$ flattening with richness/mass  when the total population was considered. If we compare their blue population LFs (which they fit with a single Schechter) and their bright part of the red population LFs (as our luminosity range spans just beyond $M_r\sim$~$-$18 in the lower $z$ bin), we find agreement as our trends with richness/mass for these galaxy populations is weak and of low significance.

\section{Discussion} 
\label{discussion}

As we discussed in Sect.~\ref{coparison_lit}, in agreement with \cite{2018Ricci} our total population $\alpha$ values increase with richness with high significance. In contrast, the values for the red and blue galaxies undergo a mild and less significant (not significant in the case of the blue galaxies) growth with richness. In the following, we try to explain this different behaviour of the total population compared to red and blue populations, taken separately.
If we look at Fig.~\ref{ztrend-2dbins} (upper right panel), as well as at Fig.~\ref{LFzrich} (upper stripe of panels), we see that the behaviour of the faint population of poor groups is dominated by the blue galaxies: the grey dots follow the blue dots at the faint end in the LF plots of Fig.~\ref{LFzrich} or, in Fig.~\ref{ztrend-2dbins}, $\alpha$ for the blue and total (grey) populations have similar value. Whereas in the rich clusters  (see Fig.~\ref{LFzrich} lower stripe of panels) the total population LFs appear to be separated from the blue ones; the faint-end slope follows the red populations more closely than the blue, especially in the higher redshift bin.
If we consider that in our sample there is a larger number of massive clusters at higher redshift ($N_{cl}$=73 in the lower right panel and $N_{cl}$=36 in the rest of the lower stripe), then we could say that the behaviour of the faint galaxies in the higher richness/mass bin is dominated by the higher redshift bin, where we found fewer faint blue galaxies (as we discuss in the following (see the lower right panel of Fig.~\ref{ztrend-2dbins}). In our study, this balance explains the total faint population trend with richness. 

\cite{2018Ricci}, who could not divide the galaxy sample by colour, claim that scenario in which the total population faint-end slope gets shallower and the $\Phi_{\star}$ parameter grows with richness requires the number of faint galaxies to be lower and the number of bright galaxies to be higher in the rich clusters compared to the groups. To explain this occurrence, they invoke luminosity enhancement by star formation in faint poor cluster galaxies; then, the number of intermediate ($\sim~ M_{\star}$) galaxies increases and the faint-end slope becomes shallower.
Moreover, we should consider the effects related to the higher errors in the photometry, which at high redshifts could bias the colour selection. 

The parameter $\Phi_{\star}$ is related to the number density of bright galaxies of magnitude $\sim M_{\star}$ and clearly increases with richness. The difference between the red and blue galaxies density, which is almost null in the groups, increases in rich clusters (see Fig.~\ref{ztrend-2dbins}, lower right panel), in a consistent way with the red/blue fraction trend on $\lambda_{\star}$ (Fig.~\ref{RBfrac}, lower panels). This implies that in massive clusters the contribution of bright red galaxies is higher than in low-mass clusters, as found also by \cite{Sarron}.

Hints about the proportion of red and blue galaxies across the different $[\lambda,z]$ bins are inferred by the area comprised between the red and blue LF counts. In Fig.~\ref{LFzrich} we see this area shrinking for increasing redshifts, indicating that if at the red bright galaxies are more abundant at low redshifts than the blue bright galaxies, at higher redshifts the blue and red counts are equal; this is  confirmed by the overlapping of the blue and red dots in the lower left panel in Fig.~\ref{ztrend-2dbins} for $z>$0.5 and by the red/blue fraction trend of Fig.~\ref{RBfrac}.
The blue faint galaxies, instead, exceed the red faint galaxies because the blue faint-end slope is steeper than the red one. This argument is supported by the $\alpha$ mild flattening  with $z$ (see Fig.~\ref{ztrend-2dbins}, upper left panel), which is also found by \cite{Sarron}. 

As is already well known, the scenario depicted by our data is compatible with massive clusters dominated by red galaxies: this is true for the bright galaxies, as we can see from the red and blue fractions trend with richness and from the $\Phi_{\star}$ trend with $\lambda_{\star}$; but the population of bright blue galaxies is also present in considerable quantity with a characteristic magnitude brighter than the red population. 
The lack of evolution in $M_{\star}$ and $\Phi_{\star}$ is compatible with a scenario in which the bright part of the LF inside $r_{200}$ is already in place at $z \sim $1 and does not significantly evolve afterwards, if not passively. 
The faint part of the red population is also not evolving, unlike the blue population, which mildly grows over time, probably accreted from the field.

\section{Conclusions} \label{conclusion}
We used the AMICO cluster catalogue from the KiDS-DR3 data \citep{Maturi} to study the LF evolution in bins of redshift and richness/mass. We selected about 4000 clusters in the redshift range [0.1,0.8] and in the mass range [$10^{13} M_{\odot}$,$10^{15} M_{\odot}$], selecting those only mildly affected by incomplete field of view and masking.   
Our sample comprises a large number of clusters providing more solid constraints on redshift evolution than other studies based on much smaller samples. It encompasses a wide range of richness and mass, including poor groups. Moreover, we computed the LFs by mapping the radial distance from the cluster centre up to $r_{200}$, and not just the central region like in many other studies, avoiding the introduction of environmental biases. 

We divided both the redshift and richness/mass ranges into 4 bins, so that we have 16 bins covering the 2D space of ($z,\lambda_{\star}$). 
The bins were subdivided so as to have a large number of galaxies per bin, ensuring solid statistical results
even if the number of clusters per bin could vary.

We separated red passive from blue star-forming galaxies by means of the colour selection prescribed by \cite{MarioBCG}, in which the effect of colour evolution of the red sequence with $z$ is taken into account.

As is already well known, our scenario is compatible with massive clusters dominated by red galaxies: this is true for the bright galaxies, as we can see from the red and blue fractions trend with richness and from the $\Phi_{\star}$ trend with $\lambda_{\star}$; but the population of bright blue galaxies is also present in considerable quantity with a characteristic magnitude brighter than the red population.

To compute LFs, we binned in magnitude and statistically subtracting the background to derive a fully model independent measure.
We stacked LF counts in each bin of ($z,\lambda_{\star}$) and for the total, red and blue galaxy populations, we fitted the stacked LF with a Schechter function and derived the parameters $\alpha$, M$_{\star}$ and $\Phi_{\star}$. Then, to study redshift evolution and richness/mass dependence, we performed a joint fit with $z$ and $\lambda_{\star}$. We found as follows:

\begin{itemize}
\item There is no faint-end redshift evolution for red and total population; the blue faint end mildly shallows, which means more faint blue galaxies at low $z$ than at high $z$ (faint blue galaxies increase).
\item The quantity M$_{\star}$ undergoes redshift passive evolution for red and total population; we do not find redshift evolution for the blue population.
\item There is no redshift evolution for density of the bright galaxies in clusters ($\Phi_{\star}$).
\item There is a mild trend in the richness for the red and blue populations in contrast to a stronger trend for the total population. 
\item There is no dependence on richness for M$_{\star}$.
\item There is a clear dependence on richness for $\Phi_{\star}$: the density of blue galaxies,
although lower, grows at a rate comparable with that of the red galaxies.
\item The values for blue M$_{\star}$ are on average brighter than for red M$_{\star}$.
\end{itemize}

These results indicate passive evolution for the LF bright part inside $r_{200}$ (already in place at $z \sim $1) and a weak increase of blue faint galaxies  probably accreted from the field. The massive clusters are dominated by red galaxies  in the LF bright part, but less significantly for increasing redshift, where the population of blue galaxies is also present in comparable quantity.
Future perspectives of this work are to use the AMICO cluster sample obtained from the KiDS-DR4 (which is in progress) to investigate how the environment and the cluster dynamical state can influence the redshift and richness/mass dependence. 
 
\begin{acknowledgements}

MS acknowledges financial contribution from contract ASI-INAF n.2017-14-H.0 and INAF `Call per interventi aggiuntivi a sostegno della ricerca di main stream di INAF'.
\\
LM acknowledges the support from the grant PRIN-MIUR 2017 WSCC32 and ASI n.2018-23-HH.0.
\end{acknowledgements}

%
%
\bibliographystyle{aa} 
\bibliography{LF-AMICO-KiDS.bib}

\begin{appendix}
\begin{table*}
    \caption{Schechter parameters derived by fitting the all, red, and blue populations stacked LFs (represented in Fig.~\ref{LFz}) in each bin of $z$ and for the whole $\lambda_{\star}$ range. The number of clusters belonging to every bin, the median values of $z_{cl}$, $\lambda_{\star}$, and log(M$_{200}$) are reported.}
    \label{table:1} 
\centering
\begin{tabular}{c c c c c c c c c}
\hline\hline                
\strut Type &range($z$) &$N_{cl}$&$\widehat{z_{cl}}$&$\widehat{\lambda_{\star}}$ &$\widehat{log M_{200}}$  & $\alpha$     & $M_{\star}$   & $\Phi_{\star}$  \\
\strut   &                           &          &                   &                           & [10$^{14}$ M$_{\odot}$] &                 & [mag]         & [mag$^{-1}$deg$^{-2}$] \\
\hline
\hline
 all  &[0.10,0.32]&  755   & 0.24                &  12  &  13.37$\pm$0.02 &-1.04$\pm$0.03 &-21.50$\pm$0.03&  2.35$\pm$0.17   \\
 red  &       &        &                 &            &                 & -               &  -            &      -  \\
 blue &       &        &                 &            &                 &-0.53$\pm$0.03 &-21.17$\pm$0.02&  1.94$\pm$0.12   \\
\hline
 all  &[0.32,0.46]&  1182  & 0.41                &  18  &  13.63$\pm$0.02 &-0.92$\pm$0.05 &-21.59$\pm$0.04&  3.91$\pm$0.39   \\
 red &                &        &                 &            &                 &-1.23$\pm$0.05 &-21.44$\pm$0.04&  1.76$\pm$0.17   \\
 blue &               &        &                 &            &                 &-0.54$\pm$0.08 &-21.51$\pm$0.07&  2.75$\pm$0.37   \\
\hline
 all  &[0.46,0.55]&  939   & 0.5                 &  22  &  13.72$\pm$0.01 &-0.87$\pm$0.07 &-21.65$\pm$0.05&  7.24$\pm$0.67   \\
 red  &               &        &                 &            &                 &-1.26$\pm$0.10 &-21.73$\pm$0.08&  3.00$\pm$0.51   \\
 blue &       &        &                 &            &                 &-0.30$\pm$0.08 &-21.42$\pm$0.05&  5.10$\pm$0.39   \\
\hline
 all  &[0.55,0.80]&  1222  & 0.63                &  29  &  13.89$\pm$0.01 &-0.63$\pm$0.11 &-21.61$\pm$0.05& 11.29$\pm$1.21   \\
 red  &               &        &                 &            &                 &-0.82$\pm$0.17 &-21.61$\pm$0.08&  6.65$\pm$1.24   \\
 blue &       &        &                 &            &                 &-0.20$\pm$0.17 &-21.41$\pm$0.07&  6.84$\pm$0.95   \\
\hline
\end{tabular} 
\end{table*}

\begin{table*}
    \caption{Schechter parameters derived by fitting the all, red, and blue populations stacked LFs (represented in Fig.~\ref{LFrichness}) in each bin of $\lambda_{\star}$ and for the whole $z$ range. The number of clusters belonging to every bin, the median values of $z_{cl}$, $\lambda_{\star}$, and log(M$_{200}$) are also tabulated.} 
     \label{table:2}
     \centering
\begin{tabular}{c c c c c c c c c}
\hline\hline                
\strut  Type  & range($\lambda_{\star}$) & $N_{cl}$ &$\widehat{z_{cl}}$ &$\widehat{\lambda_{\star}}$& $\widehat{log M_{200}}$ & $\alpha$     & $M_{\star}$   & $\Phi_{\star}$  \\
\strut   &                           &          &                   &                           & [10$^{14}$ M$_{\odot}$] &                 & [mag]         & [mag$^{-1}$deg$^{-2}$] \\
\hline\hline
 all  & [0,15]                   &  944 & 0.30              & 12                  & 13.32$\pm$0.004           & -1.05$\pm$0.07 & -21.75$\pm$0.10 &  2.97$\pm$0.40  \\
 red  &                          &      &                   &                           &                        & -1.25$\pm$0.13 & -21.60$\pm$0.16 &  1.95$\pm$0.60  \\
 blue &                          &      &                   &                           &                        & -0.53$\pm$0.02 & -21.51$\pm$0.03 &  2.31$\pm$0.09  \\
\hline 
 all  & [15,28]                  & 2040 & 0.47              & 21                  & 13.70$\pm$0.003           & -0.73$\pm$0.03 & -21.65$\pm$0.02 &  3.25$\pm$0.20  \\
 red  &                          &      &                   &                           &                        & -0.91$\pm$0.05 & -21.67$\pm$0.03 &  1.86$\pm$0.18  \\
 blue &                          &      &                   &                           &                        & -0.31$\pm$0.02 & -21.44$\pm$0.01 &  2.20$\pm$0.08  \\  
\hline 
 all  & [28,50]                  & 1005 & 0.59              & 34                  & 14.03$\pm$0.004           & -0.46$\pm$0.06 & -21.53$\pm$0.03 & 13.65$\pm$1.27  \\
 red  &                                &      &                   &                           &                  & -0.75$\pm$0.03 & -21.72$\pm$0.03 &  6.24$\pm$0.40  \\
 blue &                                &      &                   &                           &                  & -0.31$\pm$0.04 & -21.58$\pm$0.02 &  6.22$\pm$0.39  \\
\hline
 all  & [50,1000]                &  109 & 0.58              & 57$\pm$1                  & 14.43$\pm$0.02            & -0.57$\pm$0.10 & -21.52$\pm$0.08 &157.49$\pm$22.80  \\ 
 red  &                          &      &                   &                           &                        & -0.94$\pm$0.09 & -21.70$\pm$0.09 & 59.91$\pm$10.51  \\ 
 blue &                          &      &                   &                           &                        & -0.22$\pm$0.06 & -21.41$\pm$0.05 & 84.86$\pm$6.62        \\ 
\hline
\end{tabular}    
\end{table*}

\begin{table*}
    \caption{Schechter parameters derived by fitting the all/red/blue populations stacked LFs (represented in Fig.~\ref{LFzrich}) in each 2D bin of ($z$,$\lambda_{\star}$). The bin partition in the space $(z,\lambda_{\star})$ is reported. The number of clusters belonging to every bin, the median values of $z_{cl}$, of $\lambda_{\star}$ and of log(M$_{200}$) are also tabulated.}          
\label{table:3}     
\centering                         
\begin{tabular}{c c c c c c c c c c}      
\hline\hline                
\strut  Type & range($z$) & range($\lambda_{\star}$) & $N_{cl}$ & $\widehat{z_{cl}}$ & $\widehat{\lambda_{\star}}$ & $\widehat{log M_{200}}$  & $\alpha$ & $M_{\star}$ & $\Phi_{\star}$  \\        
\strut       &            &                          &          &                    &                             & [10$^{14}$ M$_{\odot}$]  &         & [mag]       & [mag$^{-1}$deg$^{-2}$] \\
\hline\hline
 all  & [0.10,0.32] & [0,15]    &  502 & 0.23      & 10 & 13.24$\pm$0.02 & -1.19$\pm$0.03 & -21.66$\pm$0.05 &    2$\pm$0   \\ 
 red  &           &           &      &             &          &                & -0.47$\pm$0.05 & -20.95$\pm$0.04 &    3$\pm$0   \\  
 blue &           &           &      &             &          &                &   -             &    -            &       -      \\
\hline
 all  & [0.32,0.46] & [0,15]    &  329 & 0.4       & 13 & 13.36$\pm$0.01 & -1.27$\pm$0.06 & -21.90$\pm$0.09 &    6$\pm$1 \\   
 red  &           &           &      &             &          &                & -0.54$\pm$0.06 & -21.52$\pm$0.06 &    7$\pm$1 \\   
 blue &           &           &      &             &          &                & -1.44$\pm$0.09 & -21.32$\pm$0.08 &    6$\pm$1 \\
\hline
 all  & [0.46,0.52] & [0,15]    &   79 & 0.48      & 13 & 13.37$\pm$0.02 & -1.55$\pm$0.07 & -22.53$\pm$0.16 &   16$\pm$4 \\  
 red  &           &           &      &             &          &                & -0.50$\pm$0.10 & -21.55$\pm$0.09 &   37$\pm$4 \\         
 blue &           &           &      &             &          &                &   -           &    -            &    -      \\
\hline
 all  & [0.52,0.80] & [0,15]    &   34 & 0.54      & 14 & 13.38$\pm$0.02 & -1.13$\pm$0.10 & -21.82$\pm$0.11 &  115$\pm$19 \\  
 red  &           &           &      &             &          &                & -0.92$\pm$0.21 & -22.17$\pm$0.30 &   50$\pm$18  \\ 
 blue &           &           &      &             &          &                & -1.53$\pm$0.23 & -22.26$\pm$0.56 &   48$\pm$37 \\
\hline
 all  & [0.10,0.32] & [15,28]   &  213 & 0.26      & 18 & 13.68$\pm$0.02 & -1.03$\pm$0.05 & -21.60$\pm$0.06 &    8$\pm$1  \\  
 red  &           &           &      &             &          &                & -0.56$\pm$0.06 & -21.27$\pm$0.05 &    7$\pm$1 \\         
 blue &           &           &      &             &          &                & -1.42$\pm$0.07 & -21.40$\pm$0.14 &    3$\pm$1 \\
\hline
 all  & [0.32,0.46] & [15,28]   &  669 & 0.41      & 19 & 13.67$\pm$0.01 & -0.83$\pm$0.04 & -21.56$\pm$0.03 &    7$\pm$0  \\  
 red  &           &           &      &             &          &                & -0.37$\pm$0.05 & -21.44$\pm$0.03 &    5$\pm$0 \\         
 blue &           &           &      &             &          &                & -1.27$\pm$0.07 & -21.50$\pm$0.05 &    3$\pm$0 \\
\hline
 all  & [0.46,0.52] & [15,28]   &  435 & 0.48      & 21 & 13.69$\pm$0.01 & -0.81$\pm$0.04 & -21.60$\pm$0.03 &   15$\pm$1 \\  
 red  &           &           &      &             &          &                & -0.28$\pm$0.06 & -21.44$\pm$0.03 &   10$\pm$1  \\  
 blue &           &           &      &             &          &                & -1.39$\pm$0.02 & -21.94$\pm$0.03 &    4$\pm$0 \\
\hline
 all  & [0.52,0.80] & [15,28]   &  723 & 0.58      & 23 & 13.72$\pm$0.01 & -0.95$\pm$0.05 & -21.80$\pm$0.03 &   10$\pm$1   \\
 red  &           &           &      &             &          &                & -0.41$\pm$0.04 & -21.51$\pm$0.02 &    7$\pm$0 \\  
 blue &           &           &      &             &          &                & -1.10$\pm$0.07 & -21.79$\pm$0.04 &    6$\pm$1 \\
\hline
 all  & [0.10,0.32] & [28,50]   &   38 & 0.28      & 33$\pm$1 & 14.13$\pm$0.04 & -0.79$\pm$0.05 & -21.36$\pm$0.06 &   95$\pm$9 \\  
 red  &           &           &      &             &          &                & -0.43$\pm$0.08 & -21.17$\pm$0.08 &   68$\pm$9  \\  
 blue &           &           &      &             &          &                & -1.39$\pm$0.05 & -21.74$\pm$0.13 &   14$\pm$3 \\
\hline
 all  & [0.32,0.46] & [28,50]   &  166 & 0.42      & 33 & 14.07$\pm$0.02 & -0.73$\pm$0.04 & -21.57$\pm$0.03 &   38$\pm$2   \\
 red  &           &           &      &             &          &                & -0.36$\pm$0.08 & -21.48$\pm$0.06 &   24$\pm$2  \\  
 blue &           &           &      &             &          &                & -1.33$\pm$0.05 & -21.91$\pm$0.07 &    8$\pm$1 \\
\hline
 all  & [0.46,0.52] & [28,50]   &  134 & 0.49      & 32 & 14.03$\pm$0.02 & -0.70$\pm$0.06 & -21.62$\pm$0.04 &   59$\pm$4 \\  
 red  &           &           &      &             &          &                & -0.21$\pm$0.06 & -21.45$\pm$0.03 &   38$\pm$2  \\  
 blue &           &           &      &             &          &                & -1.06$\pm$0.14 & -21.69$\pm$0.10 &   23$\pm$4 \\
\hline
 all  & [0.52,0.80] & [28,50]   &  667 & 0.65      & 34 & 14.01$\pm$0.01 & -0.69$\pm$0.06 & -21.73$\pm$0.04 &   19$\pm$1  \\ 
 red  &           &           &      &             &          &                & -0.24$\pm$0.06 & -21.55$\pm$0.04 &   11$\pm$1 \\   
 blue &           &           &      &             &          &                & -1.07$\pm$0.05 & -21.90$\pm$0.04 &    8$\pm$1 \\
\hline
 all  & [0.10,0.32] & [50,1000] &  962 & 0.28      & 56$\pm$1 & 14.50$\pm$0.03 & -0.57$\pm$0.14 & -20.70$\pm$0.08 & 4314$\pm$865 \\
 red  &           &           &      &             &          &                & -0.15$\pm$0.12 & -20.31$\pm$0.08 & 3160$\pm$406 \\ 
 blue &           &           &      &             &          &                &   -            &    -            &    -      \\
\hline
 all  & [0.32,0.46] & [50,1000] &   18 & 0.41      & 58$\pm$5 & 14.48$\pm$0.17 & -0.87$\pm$0.07 & -21.84$\pm$0.16 &  363$\pm$53 \\ 
 red  &           &           &      &             &          &                & -0.36$\pm$0.06 & -21.55$\pm$0.07 &  332$\pm$21 \\  
 blue &           &           &      &             &          &                & -1.23$\pm$0.14 & -21.72$\pm$0.32 &  126$\pm$49\\
\hline
 all  & [0.46,0.52] & [50,1000] &   16 & 0.485     & 56$\pm$2 & 14.44$\pm$0.06 & -0.44$\pm$0.09 & -21.47$\pm$0.08 &  802$\pm$63 \\ 
 red  &           &           &      &             &          &                & +0.20$\pm$0.15 & -21.15$\pm$0.10 &  558$\pm$22  \\ 
 blue &           &           &      &             &          &                & -1.33$\pm$0.27 & -21.95$\pm$0.30 &  150$\pm$71\\
\hline
 all  & [0.52,0.80] & [50,1000] &   73 & 0.63      & 57$\pm$1 & 14.39$\pm$0.04 & -0.15$\pm$0.06 & -21.34$\pm$0.03 &  285$\pm$9  \\ 
 red  &           &           &      &             &          &                & -0.01$\pm$0.11 & -21.37$\pm$0.05 &  131$\pm$7  \\  
 blue &           &           &      &             &          &                & -0.65$\pm$0.10 & -21.54$\pm$0.05 &  122$\pm$9 \\
\hline                                  
\end{tabular}
\end{table*}

\begin{table*}
\caption{Trend coefficients and slope significance for the redshift evolution of the Schechter parameters (plotted in Fig.~\ref{ztrend}).}          
\label{table:sigz}     
\centering                         
\begin{tabular}{c c c c}
\hline 
& intercept & $z$ slope & slope significance \\
\hline\hline
$\alpha$& & & \\
\hline     
all  & -1.25$\pm$0.16 & 0.82$\pm$0.38 & 2.1$\sigma$  \\
red  & -0.71$\pm$0.20 & 0.70$\pm$0.45 & 1.5$\sigma$  \\
blue & -1.50$\pm$0.50 & 0.78$\pm$0.97 & 0.8$\sigma$  \\
\hline\hline 
 M$_{\star}$ & & & \\
\hline 
all  & -21.44+-0.12 & -0.33+-0.26 & 1.2$\sigma$  \\
red  & -21.13+-0.24 & -0.53+-0.51 & 1.0$\sigma$  \\
blue & -21.41+-0.82 & -0.36+-1.50 & 0.2$\sigma$  \\
\hline\hline
 $\Phi_{\star}$ & & & \\
\hline 
all  & -0.06+-0.22 & 1.74+-0.48 & 3.6$\sigma$  \\
red  & -0.05+-0.20 & 1.40+-0.45 & 3.1$\sigma$  \\
blue & -0.59+-0.80 & 2.13+-1.52 & 1.4$\sigma$  \\
\hline
\end{tabular}
\end{table*}

\begin{table*}
\caption{Trend coefficients and slope significance for the  Schechter parameters dependence on richness (plotted in Fig.~\ref{richnesstrend}).}          
\label{table:sigl}     
\centering                         
\begin{tabular}{c c c c}      
\hline 
& intercept & log($\lambda_{\star}$) slope & slope significance \\
\hline\hline
$\alpha$& & & \\
\hline 
all & -1.64$\pm$0.56 & 0.67$\pm$0.40 & 1.7$\sigma$  \\
red & -0.90$\pm$0.44 & 0.38$\pm$0.30 & 1.3$\sigma$  \\
blue & -1.49$\pm$ 0.62 & 0.39$\pm$0.42 &0.9$\sigma$  \\
\hline\hline 
 M$_{\star}$ & & & \\
\hline 
all & -22.09$\pm$0.30 & 0.34$\pm$0.21 & 1.6$\sigma$  \\
red & -21.50$\pm$0.43 & 0.01$\pm$0.29 & 0.03$\sigma$  \\
blue & -21.45$\pm$0.28 & -0.16$\pm$0.19 & 0.8$\sigma$  \\
\hline\hline 
$\Phi_{\star}$ & & & \\
\hline 
all & -1.41$\pm$1.77 & 1.74$\pm$1.21 & 1.4$\sigma$  \\
red & -1.10$\pm$1.61 & 1.37$\pm$1.10 & 1.2$\sigma$  \\
blue & -1.17$\pm$1.7 & 1.38$\pm$1.14 & 1.2$\sigma$  \\
\hline                                  
\end{tabular}
\end{table*}

\begin{table*}
\caption{Coefficients resulting from the fit of the equation \ref{eq10ofRicci} to our 16 sets of Schechter parameters (plotted in Fig.~\ref{ztrend-2dbins}).}          
\label{table:sigzl}     
\centering                         
\begin{tabular}{c c c c c c}      
\hline      
& $c$ [intercept] & $a$ [$z$ slope] & $a$ significance & $b$ [log($\lambda$) slope] & $b$ significance  \\          
\hline\hline
$\alpha$& & & & & \\
\hline
all  & -2.44$\pm$0.30 & +0.36$\pm$0.93 & 0.4$\sigma$ & 1.08$\pm$0.21 & 5.2$\sigma$  \\
red  & -1.13$\pm$0.20 & +0.56$\pm$0.67 & 0.8$\sigma$ & 0.5$\pm$0.14 & 3.5$\sigma$  \\
blue & -2.38$\pm$3.64 & +2.34$\pm$1.19 & 2$\sigma$ & 0.53$\pm$2.29 & 0.2 $\sigma$  \\
\hline\hline 
 M$_{\star}$ & & & & & \\
\hline 
all  & -22.44$\pm$0.42 & +2.58$\pm$1.74 & 1.5$\sigma$ & 0.86$\pm$0.82 & 1$\sigma$  \\
red  & -21.37$\pm$0.44 & -4.67$\pm$2.21 & 2.1$\sigma$ & 0.5$\pm$0.28 & 1.8 $\sigma$  \\
blue & -21.26$\pm$0.51 & -0.68$\pm$1.36 & 0.5$\sigma$ & -0.23$\pm$0.35 & 0.6 $\sigma$  \\
\hline\hline
$\Phi_{\star}$ & & & & & \\
\hline 
all   & -2.12$\pm$0.91 & +0.46$\pm$1.75 & 0.2$\sigma$ & 2.71$\pm$0.63 & 4.3$\sigma$  \\
red   & -1.58$\pm$0.95 & -0.77$\pm$1.98 & 0.4$\sigma$ & 2.28$\pm$0.65 & 3.5$\sigma$  \\
blue  & -1.65$\pm$0.92 & +0.61$\pm$1.89 & 0.3$\sigma$ & 1.86$\pm$0.62 &3.0$\sigma$  \\
  
\hline                                  
\end{tabular}
\end{table*}

\end{appendix}

\end{document}